\documentclass[twocolumn]{aastex631}

\usepackage{amsmath}
\usepackage{tabularx}
\usepackage{threeparttable}
\usepackage{comment}
\usepackage{lipsum}

\begin{document}

\title{Why are optical coronal lines faint in active galactic nuclei?}

\correspondingauthor{Jeffrey D. McKaig}
\email{jmckaig@gmu.edu}

\author[0000-0002-0913-3729]{Jeffrey D. McKaig}
\affiliation{George Mason University, Department of Physics and Astronomy, MS3F3, 4400 University Drive, Fairfax, VA 22030, USA}

\author[0000-0003-2277-2354]{Shobita Satyapal}
\affiliation{George Mason University, Department of Physics and Astronomy, MS3F3, 4400 University Drive, Fairfax, VA 22030, USA}


\author{Ari Laor}
\affiliation{Physics Department, Technion – Israel Institute of Technology, Haifa 32000, Israel}

\author{Nicholas P. Abel}
\affiliation{College of Applied Science, University of Cincinnati, Cincinnati, OH, 45206, USA}

\author[0000-0003-3152-4328]{Sara M. Doan}
\affiliation{George Mason University, Department of Physics and Astronomy, MS3F3, 4400 University Drive, Fairfax, VA 22030, USA}

\author[0000-0001-5231-2645]{Claudio Ricci}
\affiliation{Instituto de Estudios Astrofísicos, Facultad de Ingeniería y Ciencias, Universidad Diego Portales, Av. Ejército Libertador 441, Santiago, Chile}
\affiliation{Kavli Institute for Astronomy and Astrophysics, Peking University, Beijing 100871, China}
\affiliation{George Mason University, Department of Physics and Astronomy, MS3F3, 4400 University Drive, Fairfax, VA 22030, USA}

\author[0000-0003-1051-6564]{Jenna M. Cann}
\affiliation{X-ray Astrophysics Laboratory, NASA Goddard Space Flight Center, Code 662, Greenbelt, MD 20771, USA}
\affiliation{Center for Space Sciences and Technology, University of Maryland, Baltimore County, 1000 Hilltop Circle, Baltimore, MD 21250}



\begin{abstract}
Forbidden collisionally excited optical atomic transitions from high ionization potential (IP$\geq$54.8\,eV) ions, such as Ca$^{\mathrm{4+}}$, Ne$^{\mathrm{4+}}$, Fe$^{\mathrm{6+}}$, Fe$^{\mathrm{10+}}$, Fe$^{\mathrm{13+}}$, Ar$^{\mathrm{9+}}$, and S$^{\mathrm{11+}}$, are known as optical coronal lines (CLs). The spectral energy distribution (SED) of active galactic nuclei (AGN) typically extends to hundreds of electron volts and above, which should be able to produce such highly ionized gas. However, optical CLs are often not detected in AGN. Here we use photoionization calculations with the \textsc{Cloudy} spectral synthesis code to determine possible reasons for the rarity of these optical CLs. We calculate  CL luminosities and equivalent widths from  radiation pressure confined photoionized gas slabs exposed to an AGN continuum.  We consider the role of dust, metallicity, and ionizing SED in the formation of optical CLs. We find that (1) dust reduces the strength of most CLs by $\sim$three orders of magnitude, primarily as a result of depletion of metals onto the dust grains. (2) In contrast to the CLs, the more widely observed lower IP optical lines such as [O\, III] 5007\,\AA, are less affected by depletion and some are actually enhanced in dusty gas. (3) In dustless gas many optical CLs become detectable,  and are particularly strong for a hard ionizing SED. This implies that prominent CL emission likely originate in dustless gas.  Our calculations also suggest optical CL emission is enhanced in galaxies with low mass black holes characterized by a harder radiation field and a low dust to metal ratio. The fact that optical CLs are not widely observed in the early universe with JWST may point to rapid dust formation at high redshift.
\end{abstract}

\keywords{galaxies: active -- galaxies: Evolution -- galaxies: dwarf -- line:formation -- accretion, accretion disks}

\section{Introduction}\label{sec:intro}
    
     The analysis of spectral lines is used to detect, classify, and study the physical state of  photoionized gas 
    in active galactic nuclei (AGNs; e.g., \citealp{1981PASP...93....5B, 1983ApJ...269..466K, 1987ApJS...63..295V, 1993ApJ...417...63H, 2001ApJ...556..121K, Kauffmann2003, 2006agna.book.....O}). This characterization is crucial for understanding how the supermassive black holes (SMBHs), which power AGNs, grow and evolve along with their host galaxy (e.g., \citealp{2019ARA&A..57..511K}). Collisionally excited optical lines, such as [O\,III]\,5007\,\AA\,\, and [N\,II]\,6584\,\AA,  
    are powerful tools to distinguish between AGN and star forming regions in nearby galaxies (e.g., \citealp{1981PASP...93....5B, 1987ApJS...63..295V, 2001ApJ...556..121K, Kauffmann2003}). However, these diagnostics can be modified in regions of high obscuration, excessive stellar activity, or low metallicity (e.g., \citealp{2009MNRAS.398.1165G, 2018ApJ...858...38S, 2018ApJ...861..142C}).
    
     Higher-ionization, collisionally excited optical forbidden lines, known as optical ``coronal lines" (CLs; first discovered in the solar corona) have shown great potential to uncover hidden AGNs missed by traditional optical diagnostics due to dominant star formation (e.g., \citealp{2007ApJ...663L...9S, 2009ApJ...704..439S, 2009MNRAS.398.1165G, 2015ApJ...811...26T, 2018ApJ...858...38S, 2018ApJ...861..142C, 2021ApJ...906...35S, 2021ApJ...912L...2C, 2021ApJ...922..155M, 2021ApJ...911...70B}). The ions from which optical CLs originate have ionization potentials (IPs) typically in excess of $\approx 70\,$eV and thus production by even the most energetic stellar populations is unlikely. For example, CL emission from Type II supernovae is weak and short lived (e.g., \citealp{1990AJ....100.1588B,2009ApJ...695.1334S}). Therefore, CL detection is typically considered a ``smoking gun" of AGN activity. 
    
     Optical CLs in AGNs have been studied for several decades (e.g., \citealp{1968ApJ...151..807O,1970ApJ...161..811N,1978ApJ...221..501G,1984MNRAS.211P..33P,1988AJ.....95...45A,2002MNRAS.329..309P}). In addition to their high IPs, they are also characterized by high critical densities ($\sim 10^{7}-10^{10}\,\mathrm{cm}^{-3}$) which are orders of magnitude larger than the traditionally used lower IP optical lines in the narrow line region (NLR). Thus, they are able to survive in very dense gas surrounding AGNs before being collisionally de-excited. However, optical CLs are unlikely to be associated with gas in the broad line region (BLR), as densities there are too high and would collisionally de-excite the transitions (e.g., \citealp{2014MNRAS.438..604B}). Therefore, it is often thought that they are produced in a region between the BLR and NLR: a region known as the coronal line emitting region (CLER). Observations sometimes show that the CLER is indeed more compact than that of the stronger lower ionization optical lines \citep[e.g.,][]{1988AJ.....95.1695V, 2010MNRAS.405.1315M}. For example, VLT/GRAVITY observations of NGC 3783 \citep{2021A&A...648A.117G} placed the [Ca\,VIII] emission at the inner edge of the NLR at $\sim$0.4\,pc. Measurements of the full width at half max (FWHM) of CLs have shown values up to  $\sim 1000-2000\,\mathrm{km\,s}^{-1}$ (e.g., \citealp{1984ApJ...286..171D, 1986ApJ...301..727D, 1988AJ.....95.1695V, 1988AJ.....95...45A, 1991A&A...250...57A, 2006ApJ...653.1098R}), consistent with the proposed intermediate nature of the CLER in some objects. However, recent surveys have shown optical CL FWHMs similar to that of non-CL narrow lines (e.g., \citealp{2023ApJS..265...21R}, Matzko et al. submitted, Doan et submitted), which suggests an origin which extends further out and overlaps the NLR.  

     What physical mechanism can produce a compact CLER? \cite{1995ApJ...450..628P} proposed that the inner edge of the torus is driven off due to an X-ray heated wind. Such a wind is heated to the Compton temperature which sputters the dust grains, creating a ``natural place" for iron CLs (e.g., [Fe\,X]\,6374\,\AA). \cite{1998ApJ...497L...9M} found that type\,1 AGNs have excess [Fe\,VII] 6087\,\AA\,\,detections compared to type\,2 AGNs, implying the CL emission is compact and also obscured in type\,2s. A compact CLER is also proposed for extremely rare objects known as coronal line forest (CLiF) AGN, characterized by a rich spectrum of optical CLs \citep{2015MNRAS.451L..11R, 2015MNRAS.448.2900R, 2016ApJ...824...34G, CC2021}. However, CL emission has also been shown to extend up to the $\sim 10$\,kpc scale from the accreting system of many AGNs (e.g., \citealp{2005MNRAS.364L..28P, 2018ApJ...858...48M, 2018MNRAS.481L.105M, 2021ApJ...920...62N,2019BAAA...61..186D, 2023ApJ...945..127N}; Matzko et al.\,\,submitted). The natural explanation of such extended CL emission is either AGN photoionization or shocks, with the relative contribution being debated in the literature (e.g., \citealp{laor1998, 1998MNRAS.298.1035T, 2005MNRAS.364L..28P, 2020ApJ...895L...9R, 2023arXiv230713263N}). While shock models are consistent with the CL emission in some AGNs \citep[e.g.,][]{2005MNRAS.364L..28P, 2011A&A...533A..63E, 2020ApJ...895L...9R, 2021ApJ...922..155M}, AGN photoionization models can also produce strong CL emission up to 10\,kpc from the ionizing source (e.g., \citealp{2014MNRAS.438..901S}). In this work we consider photoionization models for the production of CL emission, and discuss why shocks are an unlikely source for CL in AGN.

     What can optical CLs tell us about the illuminating spectral energy distribution (SED)? The IPs of the optical CL emitting species is typically well above $\sim$4\,Ryd, e.g. $\sim$0.5\,keV for S$^{11+}$ (which produces the [S\,XII]\,7611\,\AA\ line). The extreme UV region of the SED cannot be directly observed due to Galactic and intrinsic absorption (e.g., \citealp{1996A&A...315L.109M,2000ApJ...536..710A}), and may be tied to fundamental black hole parameters such as mass, Eddington ratio, and spin. The IPs of optical CLs provide a useful proxy for this energy range and may allow to constrain accretion disk models along with the fundamental black hole parameters (e.g., \citealp{2018ApJ...861..142C, 2022MNRAS.510.1010P}, Bierschenk et al. submitted). CLs have been used to uncover low mass black holes in dwarf galaxies, with no signs of AGN activity otherwise (e.g., \citealp{2021ApJ...906...35S, 2021ApJ...922..155M, 2023ApJ...946L..38R}), and may have the potential to uncover the yet elusive intermediate mass black hole (IMBH; e.g., \citealp{2018ApJ...861..142C, 2021ApJ...912L...2C, 2021ApJ...922..155M, 2022ApJ...936..140R,2023ApJ...946L..38R}). However, some studies have called into question the presence of a connection between the observed continuum emission and fundamental parameters such as black hole mass and Eddington ratio (e.g., \citealp{2014MNRAS.438.3024L, 2022arXiv221011977M}).

    Systematic surveys have studied the incidence of CLs in both AGNs (local and high redshift; e.g.,  \citealp{2002ApJ...579..214R, 2006A&A...457...61R, MS_2011, 2011ApJ...743..100R, 2017MNRAS.467..540L, 2018ApJ...858...48M, 2023ApJ...948..112C}) and in the general galaxy population (e.g., \citealp{2009MNRAS.397..172G, 2021ApJ...920...62N, 2023ApJ...945..127N}). Although CLs are seen in the majority of nearby bright Seyferts, recent large optical spectroscopic surveys using the Sloan Digitial Sky Survey (SDSS) have shown that detections of CLs are generally rare in both type\,1 and type\,2 AGNs (e.g., \citealp{2009MNRAS.397..172G, 2022ApJ...936..140R, 2023ApJ...945..127N}, Doan et al. submited).  Specifically,  Doan et al.\,\,found only 4.5$\%$ of type\,1 AGNs from the SDSS DR7 release,
    display at least one optical CL with an equivalent width above $1\,$\AA. Even more rare are CLiF AGN in SDSS, defined by strong optical CLs relative to the H recombination lines. CLiF AGN show flux ratios such as $F(\text{[Fe\,VII] 6087\,\AA}) / F(\text{H}\alpha)>0.25$ and $F(\text{[Ne\,V] 3425\,\AA}) / F(\text{H}\alpha)>0.2$ (Rose et al 2015, section 2 there). These objects generally show CL equivalent widths (EWs) in the range $\sim\,1-10\,$\AA\ (no CLiF AGN show CL EWs$>10\,$\AA). Given that essentially all AGNs produce a copious supply of high energy photons with energies sufficient to produce CL emitting ions, why are the optical CLs faint in most AGNs?

    Our main goal in this work is to study the strength of optical CLs as a function of the gas dust content, the gas metallicity, and the ionizing SED shape, to understand what drives their faintness relative to the strong lower IP narrow optical lines. We use the spectral synthesis code \textsc{Cloudy} \citep{2017RMxAA..53..385F}, including the radiation pressure compression 
    (RPC) effect, to calculate luminosities and EWs of various optical lines, coronal and non coronal. We outline the physical model used in $\S$\ref{sec:modeling}, present our results in $\S$\ref{sec:results}, a discussion in $\S$\ref{sec:discussion}, and our conclusions is in $\S$\ref{sec:conclusions}.
    
\begin{table*}\label{linetable}
\caption{Table of 12 optical CLs to be considered in this study along with their critical densities, IPs, and transitions.}
\begin{tabular}{lcccc}
\hline
\hline
Line          & Wavelength (\AA)$^1$ & Critical Density (cm$^{-3}$) & Ionization Potential (eV)$^{2}$ & Transition                                                                             \\ \hline
{[}Fe\,V{]}   & 3891.28                         & $1.61\times 10^{8}$          & 54.84                     & $^{5}$D$_{4}-^{3}$F$2_{4}$                                                        \\
{[}Ca\,V{]}   & 5309.11                         & $6.36\times 10^{7}$          & 67.10                     & $^{3}$P$_{2}-^{1}$D$_{2}$                                                \\
{[}Fe\,VI{]}  & 5335.18                         & $6.32\times 10^{6}$          & 75.00                     & $^{4}$F$_{3/2}-^{4}$F$_{1/2}$                                            \\
{[}Fe\,VI{]}  & 5176.04                         & $3.29\times 10^{7}$          & 
75.00                     & $^{4}$F$_{9/2}-^{2}$G$_{9/2}$
\\
{[}Ne\,V{]}   & 3425.88                         & $1.90\times 10^{7}$          & 97.11                     & $^{2}$P$_{2}-^{1}$D$_{2}$                                                \\
{[}Fe\,VII{]} & 6087.00                         & $4.46\times 10^{7}$          & 99.00                     & $^{3}$F$_{3}-^{1}$D$_{2}$                                                              \\
{[}Fe\,VII{]} & 4893.37                         & $3.09\times 10^{6}$          & 99.00                     & $^{3}$F$_{2}-^{3}$P$_{1}$                                                              \\
{[}Fe\,X{]}   & 6374.51                         & $4.45\times 10^{8}$          & 235.04                    & $^{2}$P$_{3/2}^{0}-^{2}$P$_{1/2}^{0}$                                                  \\
{[}Fe\,XI{]}  & 7891.80                         & $6.39\times 10^{8}$          & 262.10                    & $^{3}$P$_{2}-^{3}$P$_{1}$                                                              \\
{[}Fe\,XIV{]} & 5302.86                         & $3.99\times 10^{8}$          & 361.00                    & $^{2}$P$_{1/2}^{0}-^{2}$P$_{3/2}^{0}$                                                  \\
{[}Ar\,X{]}   & 5335.27                         & $2.36\times 10^{9}$          & 422.60                    & $^{2}$P$_{3/2}^{0}-^{2}$P$_{1/2}^{0}$                                                  \\
{[}S\,XII{]}  & 7611.00                         & $7.09\times 10^{9}$          & 504.78                    & $^{2}$P$_{1/2}^{0}-^{2}$P$_{3/2}^{0}$                                                  \\ \hline
\end{tabular}
\begin{tablenotes}
	\item[0] \textsuperscript{1}Wavelengths taken from: \url{https://physics.nist.gov/PhysRefData/ASD/lines_form.html}.
	\item[1] \textsuperscript{2}Ionization potential taken from: \url{https://physics.nist.gov/PhysRefData/ASD/ionEnergy.html}.
\end{tablenotes}
\end{table*}

\section{Modeling}\label{sec:modeling}   

     In this work, we follow the RPC modeling approach from \cite{2014MNRAS.438..901S} and \cite{2014MNRAS.438..604B} which takes into account the radiation pressure force and the resulting compression of the ionized gas. The RPC effect provides the mechanism which sets the gas density profile and its ionization state, in contrast with the non RPC photoionization models, where 
     the gas density is assumed to be constant with a value which is a free parameter. Thus, the RPC gas ionization structure and the line emission are uniquely set by the distance of the gas from the ionizing source, in contrast with the constant density models, such as the Local Optimally Cloud (LOC) models (\citealp{1995ApJ...455L.119B, 1997ApJ...487..122F}), where the line emission depends on the assumed gas density.

     The RPC effect leads to a density profile which increases with depth into the photoionized gas, as the gas gets compressed by the incident radiation. The surface layer, which is lower density and highly ionized, and can produce CL emission, 
     while the inner higher density layers are less ionized, and produce lower ionization lines (see $\S$\ref{sec:RPC}).
     The RPC effect implies a much wider range of ionization states in the photoionized gas, compared to the constant density models. Photoionization modeling of the NLR, which include the RPC effect, were carried out in various studies of AGN, in particular \cite{Binette1997, Dopita2002, Groves2004a, Groves2004b, Davies2016}, and  \cite{2023ApJ...954..175Z}, where it was shown to reproduce the general properties of AGN photoionized gas such as the observed ionization parameter (which is set by the RPC solution) and various line ratios. It is important to note that the RPC effect is not a model assumption, but rather an inevitable physical effect when radiation strikes gas. 

     The RPC effect removes the major source of uncertainty in the predicted line emission properties at a given radius, which is the gas density, or equivalently the gas ionization parameter. The remaining uncertainties in the predicted line emission at a given radius are the ionizing SED shape, the gas metallicity, and the presence of dust, which are generally but not always (as we show below), only of second order importance. Below we explore how these remaining free parameters in the RPC model affect the CL intensities, and what may explain the CL general faintness in AGN.

     The RPC effect also provides a confining medium for the photoionized gas, which would otherwise 
     evaporate on short timescales. The RPC effect will not be significant if the photoionized gas is
     confined by another mechanism, such as magnetic confinement, which happens to lead to gas pressure 
     well above the incident radiation pressure. If this situation applies, the photoionized gas will
     be denser and less ionized than expected from the RPC effect. In highly photoionized gas, such as the gas producing the CLs, the radiation pressure is necessarily well above the gas pressure, and the RPC
     effect will dominate.
    
    \subsection{Radiation Pressure Compression}\label{sec:RPC}
        
        When a cloud of gas is irradiated by an external source, such as that of an AGN, photons transfer both energy and momentum into the gas. Assuming the cloud is in a Keplerian orbit and hydrostatic equilibrium, the interplay between radiation pressure ($P_{\mathrm{rad}}$) and gas pressure ($P_{\mathrm{gas}}$) will dominate the internal cloud gas dynamics as long as there exist no other significant sources of internal pressure such as magnetic or turbulent pressure. In particular, if the incident radiation pressure dominates over the local gas pressure at the illuminated face, absorption of the incident radiation field will lead to an inevitable compression of the photoionized layer. Thus, the gas density, temperature, and pressure inside the cloud are no longer free parameters, but a function of the incident radiation pressure. As the ionization front is approached, the incident radiation 
        absorption increases and the gas density continues to increase, until the radiation is fully absorbed and the local gas pressure reaches the incident radiation pressure. 

        Assuming radiation and gas pressure are the sole forces acting on the cloud, an estimate of the density at the ionization front can be reached \citep[see][]{2014MNRAS.438..901S}. A hydrostatic solution implies that the radiative force exerted by the absorption of the ionizing radiation is balanced by the counter force produced by the gas pressure gradient, that is by the increase in $P_{\mathrm{gas}}$ into the gas cloud:
        \begin{equation}\label{dPgdz}
            \begin{split}
                \frac{dP_{\mathrm{gas}}}{dz} &= -\frac{dP_{\mathrm{rad}}}{dz} \\
                &= \beta\frac{F_{\mathrm{rad}}}{c} n_{\mathrm{H}}\Bar{\sigma}
            \end{split}
        \end{equation}
        where $F_{\mathrm{rad}}$ is the flux of ionizing photons (roughly constant in the optically thin layer of the slab), $n_{\mathrm{H}}$ is the number density of Hydrogen,  $\Bar{\sigma}$ is the mean total opacity per H nucleus, and $\beta$ is a correction factor for the absorption of non-ionizing photons, where $\beta\approx 1$ and $\approx 2$ in dustless and dusty gas (for convenience we assume below $\beta= 2$), respectively. 
        This relation implies that the drop in radiation pressure is associated with an equal rise in the gas
        pressure, so their total pressure remains constant (assuming a negligible initial gas pressure). 
        That is 
         \begin{equation}\label{PplusPeqC}
                P_{\mathrm{gas}}(z) + \beta P_{\mathrm{rad}}(z) = \mathrm{constant}.
        \end{equation}
         At the ionization front where most of the incident radiation is absorbed, $T\approx 10^{4}\,$K \citep{1999agnc.book.....K} and the final gas pressure ($P_{\mathrm{gas,f}}$) will equal the incident radiation pressure, that is 
         \begin{equation}\label{eqU}
                   P_{\mathrm{gas, f}} = 2.3n_{\mathrm{f}}k_{\mathrm{B}}T =  \beta P_{\mathrm{rad}}= \beta n_{\gamma} \langle h\nu \rangle  
        \end{equation} 
         where $k_{\mathrm{B}}$ is Boltzmann's constant, $n_{\gamma}$ is the ionizing photon density, and $\langle h\nu \rangle$ their mean energy. The ionization parameter at the ionization front is then 
        \begin{equation}\label{eqU1}
                  U_{\mathrm{f}}=\frac{n_{\gamma}}{n_{\mathrm{f}}}  = \frac{2.3k_{\mathrm{B}}T}{\beta\langle h\nu \rangle} \simeq 0.03 \ ,
        \end{equation} 
        using values typical for luminous AGN ($k_{\mathrm{B}}T\simeq 1$\,eV, and $\langle h\nu \rangle\simeq 25$\,eV). Thus, RPC implies that $U_{\mathrm{f}}$ is set by the $\langle h\nu \rangle$ which is characteristic of the incident SED. This value is generally $\sim$ 25\,eV for local AGN, leading to $U_{\mathrm{f}}\sim 10^{-2}$, which is generally observed (e.g., \citealp{F&N1983}). However, there are classes of objects such as low luminosity AGN (LLAGN) and low ionization nuclear emission line regions (LINERs) that are observed to have $U_{\mathrm{f}}\lesssim 10^{-3}$ (e.g., \citealp{F&N1983, Groves2004b, Molina2018}). These objects may have  $\langle h\nu \rangle\gg 25$\,eV, which will lower $U_{\mathrm{f}}$ potentially to this value. Additionally, the host stellar ionization can dominate on the NLR scales in these objects, also lowering the value of $U_{\mathrm{f}}$.          
 
        Following eq.~\ref{PplusPeqC},  the density at the ionization front can then be found as 
        \begin{equation}\label{eqdensity}
            \begin{split}
                P_{\mathrm{gas, f}} &= 2.3n_{\mathrm{f}}k_{\mathrm{B}}T = \frac{\beta L_{\mathrm{ion}}}{4\pi r^{2}c}\\
                &\Rightarrow n_{\mathrm{f}} = 1.76\times 10^{8}\left(\frac{L_{\mathrm{ion}}}{10^{45}\,\mathrm{erg\,s}^{-1}}\right)\left(\frac{r}{\mathrm{pc}}\right)^{-2}\mathrm{\,cm}^{-3}
            \end{split}
        \end{equation} 
        
        for dusty gas where $r$ is the distance in parsecs from the illuminated face of the cloud to the ionizing radiation source, and $L_{\mathrm{ion}}$ is the ionizing luminosity of the incident radiation in $\,\mathrm{erg}\,\mathrm{s}^{-1}$ \citep{2014MNRAS.438..901S}. In dustless gas, $n_{\mathrm{f}}$ will be smaller by a factor of $\beta$. The density increases as the ionizing radiation is absorbed, and complete absorption of the radiation occurs on the length scale, $l_{\mathrm{p}}$, given by (for dusty gas)
        \begin{equation}\label{eqlp}
            l_{\mathrm{p}} = 5.70\times 10^{14}\left(\frac{\Bar{\sigma}}{10^{-21}\,\mathrm{cm}^{2}}\right)^{-1}\left(\frac{L_{\mathrm{ion}}}{10^{45}\,\mathrm{erg\,s}^{-1}}\right)^{-1}\left(\frac{r}{\mathrm{pc}}\right)^{2}\,\mathrm{cm}
        \end{equation}
        where $\Bar{\sigma}$ is the mean extreme UV absorption cross section for dusty gas \citep{1993ApJ...402..441L}. Therefore we generally get that $l_{\mathrm{p}}/r\ll 1$ at all relevant distances. For example, dusty gas exposed to an ionizing luminosity of $L_{\mathrm{ion}} = 10^{45}\,\mathrm{erg\,s}^{-1}$, maintains $l_{\mathrm{p}} / r < 1$ for $r \lesssim 2\times 10^{22}\,$cm or $r \lesssim 5.4\,$kpc. Therefore, RPC clouds in the NLR of AGN can be approximated as thin, plane parallel slabs, most likely thin photoionized surface layers of much larger gas clouds. 

        As can be seen in equations \ref{eqdensity} and \ref{eqlp}, the radial structure of clouds confined by radiation pressure depends only on the ratio $L_{\mathrm{ion}} / r^{2}$ and the gas properties given by $\Bar{\sigma}$, with no other free parameters. The gas density and temperature, which give the required gas pressure, are calculated numerically under the condition of photoionization equilibrium. 
        Furthermore, the gas density at the ionization front, where most of the ionizing radiation is already absorbed, and $T\sim 10^4$~K, is set only by $L_{\mathrm{ion}} / r^{2}$. 
        The RPC model has been shown to reproduce many of the observed AGN emission line properties, such as the $r^{-2}$ density relation in the NLR of AGN \citep{2014MNRAS.438..901S}, the relative intensity of lines in the BLR \citep{2014MNRAS.438..604B}, the ionization distribution of AGN outflows \citep{2014MNRAS.445.3011S}; as well as properties of quasar absorption line outflows \citep{2014MNRAS.445.3025B}, the free-free absorption and emission signatures in AGN \citep{2021MNRAS.508..680B}, and the X-ray differential emmission measure in the NLR \citep{2019MNRAS.485..416B}.      
        
    \subsection{Numerical Calculations}\label{sec:Cloudy}

        We use the spectral synthesis code \textsc{Cloudy} \citep{2017RMxAA..53..385F} v17.03 to carry out numerical calculations for a grid of RPC slabs at various distances from the ionizing source. We place these slabs at distances ranging from $10^{17}$ to $10^{21}$\,cm in steps of 0.2 dex, with $10^{17}$\,cm being the outer radius of the BLR due to dust sublimation for an AGN luminosity of $L_{\mathrm{bol}} = 10^{45}\,$erg\,s$^{-1}$ (\citealp{1987ApJ...320..537B, 1993ApJ...402..441L, 2008ApJ...685..160N, 2018MNRAS.474.1970B}). It will be seen below that a stopping distance of $10^{21}\,$cm is sufficient to show the behavior of the optical CLs into the NLR. We consider different dust to gas ratios and gas metallicities to test their effect on the emission of optical CLs. To this end, we run separate calculations for dusty and dustless gas, using the \textsc{Cloudy} built-in depleted ISM abundance set with ISM grains for dusty gas, and solar abundances from \cite{2010Ap&SS.328..179G} for dustless gas. We consider two different values of metallicity, $Z$, either $Z = 2Z_{\mathrm{ISM}}$, or $0.2Z_{\mathrm{ISM}}$ when grains are present and $Z = 2Z_{\odot}$, or $0.2Z_{\odot}$ when grains are absent. We explore supersolar metallicity because higher metallicities are likely to be common in the NLR of optically-identified AGNs \citep[e.g.][]{2013MNRAS.431..836S}. All metals are scaled linearly with $Z$ except Helium and Nitrogen, which are scaled according to \cite{2004IAUS..222..263G}. We assume a constant dust-to-metals (DTM) ratio when scaling the metallicity and therefore scale the grain abundances by the same factor as the metal abundance. This is equivalent to scaling the dust-to-gas (DTG) ratio linearly with $Z$. This linear relationship between the DTG ratio and $Z$ seems to break down when considering a large dynamic range in $Z$, but generally holds when $Z\geq0.2Z_{\odot}$ \citep{2013A&A...557A..95R, 2015MNRAS.449.3274F, 2019A&A...623A...5D, Galliano2021}. The line luminosity ($L_{\mathrm{line}}$) scales linearly with the covering factor of the gas (CF), and \cite{2012MNRAS.426.2703S} found in a sample of broad H$\alpha$ selected type\,1 AGN from SDSS, an $L_{\mathrm{bol}} = 10^{45}\,$erg\,s$^{-1}$ corresponded to a CF of $\sim$0.04 when neglecting dust absorption. However, when dust is present, only about $1/9$ of ionizing photons are absorbed by the gas \citep{2014MNRAS.438..901S}. Therefore, we choose a CF in our calculations of $\mathrm{CF} = 9\times0.04\sim 0.3$ or $30\%$. 

        We consider the AGN SED used in \cite{2014MNRAS.438..901S} and \cite{2014MNRAS.438..604B} described as follows. The SED shape is given by the piecewise function: 
        \begin{equation}\label{eqfnu}
            f\propto
            \begin{cases}
                \nu^{\alpha_{\mathrm{UV}}}e^{-h\nu / k_{\mathrm{B}}T_{\mathrm{BB}}}e^{-k_{\mathrm{B}}T_{\mathrm{IR}} / h\nu} & 1\,\mu\text{m}-1\text{\,Ryd} \\
                \nu^{\alpha_{\mathrm{ion}}} & 1\,\text{Ryd}-1\text{\,keV} \\
                \nu^{\alpha_{\mathrm{X}}} & 1\,\text{keV}-100\text{\,keV} 
            \end{cases}
        \end{equation}
        where $\alpha_{\mathrm{UV}} = -0.5$, $k_{\mathrm{B}}T_{\mathrm{BB}} = 13\,$eV, $k_{\mathrm{B}}T_{\mathrm{IR}} = 0.1\,$eV, and $\alpha_{\mathrm{X}} = -1$ (all values from \citealp{2014MNRAS.438..604B}). The value $f$ is assumed to be zero longwards of $1\,\mu$m and above $100\,$keV. We consider three different values of $\alpha_{\mathrm{ion}}$: $-2.0$, $-1.6$, and $-1.2$, to determine the effect of a changing ionizing slope on the optical CL production. This range of $\alpha_{\mathrm{ion}}$ matches the observed range in AGN over a wide range
        of luminosities \citep[e.g. the compilation in figure 8 in][]{Lusso2015}.
        These values correspond to an $\alpha_{\mathrm{ox}} = -1.16, -1.45,$ and $-$1.74 for each value of $\alpha_{\mathrm{ion}}$ respectively. This range of $\alpha_{\mathrm{ox}}$ values covers the observed 
        $\alpha_{\mathrm{ox}}$ versus $L_{\mathrm{bol}}$ relation in type 1 AGN at $L_{\mathrm{bol}} = 10^{43}-10^{47}\,$erg\,s$^{-1}$ \citep{Steffen2006,Just2007}. We note in passing that the 
        $\alpha_{\mathrm{ox}}$ versus $L_{\mathrm{bol}}$ relation appears to extend to LLAGN at
        $L_{\mathrm{bol}} < 10^{40}\,$erg\,s$^{-1}$, where $\alpha_{\mathrm{ox}} > -1$ \citep{Bianchi2022}. 
        Therefore, the ionizing SEDs explored here apply strictly only to AGN at $L_{\mathrm{bol}} > 10^{43}\,$erg\,s$^{-1}$. 
        We normalize equation\,\ref{eqfnu} to have a bolometric luminosity of $L_{\mathrm{bol}} = 10^{45}\,$erg\,s$^{-1}$. There do exist many accretion disk models which tie the SED back to the fundamental black hole and accretion properties such as mass, Eddington ratio, and spin \citep[e.g.,][]{2012MNRAS.420.1848D, 2023MNRAS.520.2781K}. There have also been \textsc{Cloudy} models which have taken these SEDs into account to explain high energy line formation \citep[e.g.,][]{2018ApJ...861..142C, 2019ApJ...870L...2C}. However, our goal in this paper is to determine what physical conditions create optical CLs, and to explain their relative faintness in the majority of AGNs. It is not our intent in this work to model the observed CL emission in individual AGNs, or to explore their relationship with fundamental black hole properties.  Therefore, we chose the observed SED shown in eq.\,\ref{eqfnu}.   
        
        We choose an ionization parameter at the cloud's face, $U_{\mathrm{i}}$, defined as
        \begin{equation}\label{eqU2}
            U_{\mathrm{i}} = \frac{\phi_{\mathrm{H}}}{cn_{\mathrm{H}, \mathrm{i}}}
        \end{equation}
        where $\phi_{\mathrm{H}}$ is the Hydrogen ionizing flux from the SED, $c$ is the speed of light, and $n_{\mathrm{H}, i}$ is the number density of Hydrogen at the illuminated face, of $U_{\mathrm{i}} = 1000$. The gas is compressed by the radiation absorbed inside the cloud, as a result $U$ decreases reaching $U\simeq 0.01$ at the ionization front (eq.~\ref{eqU1}), independent of the value of $U_{\mathrm{i}}$,
        as long as $U_{\mathrm{i}}\gg 0.01$.   
        The value of $U_{\mathrm{i}}$, along with $r$ and $L_{\mathrm{bol}}$, sets the density at the illuminated face of each cloud via eq.\,\ref{eqU}. The choice of $U_{\mathrm{i}} = 1000$ is made for two reasons: (1) it ensures 
        $P_{\mathrm{rad}}\gg P_{\mathrm{gas}}$ at the cloud face, so significant gas compression occurs \citep{2014MNRAS.438..901S} and (2) the gas is fully ionized at $U_{\mathrm{i}} = 1000$, and no lines are produced in that layer. The solution will remain the same if we choose $U_{\mathrm{i}}>1000$, as higher values will just produce a thicker layer of fully ionized gas with no line emission, and it will lead to the same slab structure for the compressed gas where the lines are produced.  Note also that generally $U_{\mathrm{i}}\gg 1$ is required to include the higher ionization layers  where the CLs are produced.   

        We stop each calculation at the Hydrogen ionization front where the H$^{+}$ fraction drops below 0.1. Here the electron temperature will be $T\leq 10^{4}\,$K and there will be little contribution to optical line emission.

        \begin{figure*}
            \centering
            \includegraphics[scale = 0.437]{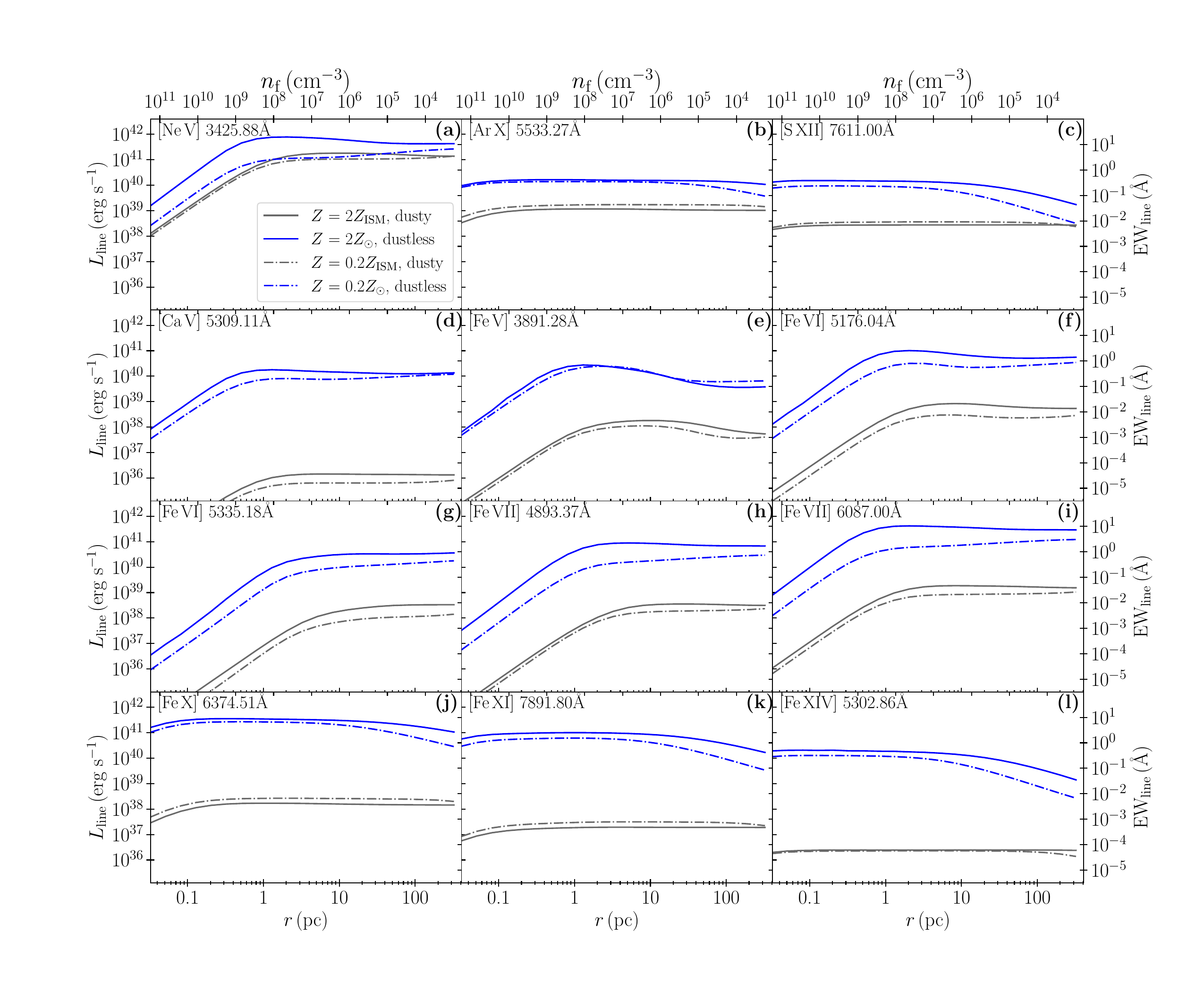}
            \caption{The \textsc{Cloudy} RPC results for the luminosity and EW (at 5100\,\AA) for the various optical CLs (Table\,\ref{linetable}) as a function of $r$. The models are for dusty and dustless gas, at low and high $Z$. All models assume $\alpha_{\mathrm{ion}} = -1.6$ and $\mathrm{CF} = 0.3$. The top axis denotes the final density $n_{\mathrm{f}}$ of a RPC 
            slab placed at $r$ (see eq.\ref{eqdensity}). Each line is collisionaly suppressed at a small enough $r$,  where $n_{\mathrm{f}} > n_{crit}$ 
            for that line. Note the overall dramatic suppression of the lines in dusty gas,  in particular for the 
            iron and calcium lines, which are heavily depleted into grains. The [Ne\,V] 3425.88\AA\,\,line is predicted to 
            be the strongest optical CL, and is likely detectable (EW$>1$\AA) also in dusty gas. The next strongest
            line is [Fe\,VII] 6087.00\AA, but it is detectable only in dustless gas.}  
            \label{fig1}
        \end{figure*}

        In \textsc{Cloudy}, constant pressure is achieved using the \texttt{constant pressure} command. However, in dustless calculations including line radiation pressure cause stability problems. Therefore, line radiation pressure is turned off for all calculations with the \texttt{no line transfer} command. As shown in \cite{2014MNRAS.438..604B} Figures 1, 2, and B1, this could change some line EWs by a factor of upto $\sim$2.

        Finally, our calculations take advantage of \textsc{Cloudy}'s `luminosity mode' which outputs total line luminosities at the end of each calculation. This allows us to examine line emission as a function of distance from the AGN source. In the Coronal Line Spectroscopic Survey, \cite{2022ApJ...936..140R} searched the entire SDSS catalog for 20 optical CLs (see \citealp{2022ApJ...936..140R} Table\,1). Many of these lines come from similar upper levels of the same ion with only slightly differing critical densities (e.g., [Ne\,V] 3425.88\,\AA\,\, and [Ne\,V] 3345.82\,\AA). Thus, we expect the excitation of these lines to be similar. Using this reasoning, we choose to only consider the 12 independent optical CLs shown in Table \ref{linetable} here. In order to make a statement on the observability of these lines for a given calculation, we also report EWs using the incident continuum density at 5100\,\AA. An EW $\geq 0.1\,$\AA\,\, may be observable with high enough S/N. However, EWs are only valid when considering observations of a type\,1 source where the illuminating continuum is also observed. When comparing with Type\,2 observations with potential host obscuration, line ratios with a more prominent AGN-driven line, such as [O\, III] 5007\,\AA\,\, is a more useful indicator of the line strength.

\section{Results}\label{sec:results}

    \subsection{The optical CL}\label{sec:CLresults}

        Figure \ref{fig1} presents the \textsc{Cloudy} results for the luminosities and EWs of the CLs in Table~\ref{linetable} as a function of the distance from the ionizing source (0.03$-$300~pc). The results are presented for the low ($0.2Z_{\odot}$) and high metallicity ($2Z_{\odot}$) cases, and both metallicity cases are presented for
        either dusty and dustless gas. The AGN luminosity is $L_{\mathrm{bol}} = 10^{45}\,$erg\,s$^{-1}$, and the gas absorbs 30\% of the ionizing radiation (CF=30\%).  The figure also designates the final density of a dusty RPC slab placed at a given $r$ (top axes), which scales as $r^{-2}$ (dustless gas will have a lower density by a factor of $\sim$2, see eq.~\ref{eqdensity}).

        Overall, the strongest CL lines are [Ne\,V] 3425.88\AA\ and 
        [Fe\,VII] 6087.00\AA,  which can reach EW $\geq 10$\AA\ for dustless high Z gas. The set of lines which are weaker but can reach EW $\geq 1$\AA\ are
        [Fe\,VI] 5176.04\AA, [Fe\,VII] 4839.37\AA, [Fe\,X] 6347.51\AA, and [Fe\,XI] 7891.80\AA. All other lines can reach EW $\geq 0.1$\AA\ in dustles gas and thus may become detectable in high S/N spectra.
        
        We see that the relatively lower ionization lines in Fig.\ref{fig1} ([Ne\,V] 3425.88\AA, [Ca\,V] 5309.11\AA, [Fe\,V] 3891.28 \AA, [Fe\,VI] 5176.04, 5335.18 \AA, and [Fe\,VII] 4839.37, 6087.00\AA)
        increase significantly in strength with distance, with a maximum at $r\sim 1$~pc followed by only a slight drop at larger $r$. This dependence is the critical density ($n_{\mathrm{crit}}$) effect. At small radii where $n_{\mathrm{f}} > n_{\mathrm{crit}}$, transitions from upper to lower levels are collisionally de-excited, so  $L_{\mathrm{line}}\propto n_{\mathrm{crit}}/n_{\mathrm{f}}$, or $L_{\mathrm{line}}\propto r^{2}$. At larger radii where $n_{\mathrm{f}}< n_{\mathrm{crit}}$, the radiative transitions to lower levels dominate, the lines reach their peak luminosity, with a slight drop at larger $r$. The higher ionization lines in Fig.\ref{fig1} ([Ar\,X] 5533.27AA\, [S\,XII] 7611.00 \AA, [Fe\,X] 6347.51\AA, [Fe\,XI] 7891.80\AA, and [Fe\,XIV] 5302.86\AA)
        show only a slight drop at the smallest radii, as these transitions are all characterized by higher critical densities, so collisional de-excitations do not become dominant even at the smallest radii probed here. Therefore, stratification with $r$ is expected with higher IP lines peaking closer to the center. The highest ionization lines generally peak at $\sim$0.1\,pc, while lower IP lines peak at 1$-$10\,pc, or possibly further out \citep[e.g.,][]{2014MNRAS.438..901S}.

        \begin{figure*}
            \centering
            \includegraphics[scale = 0.43]{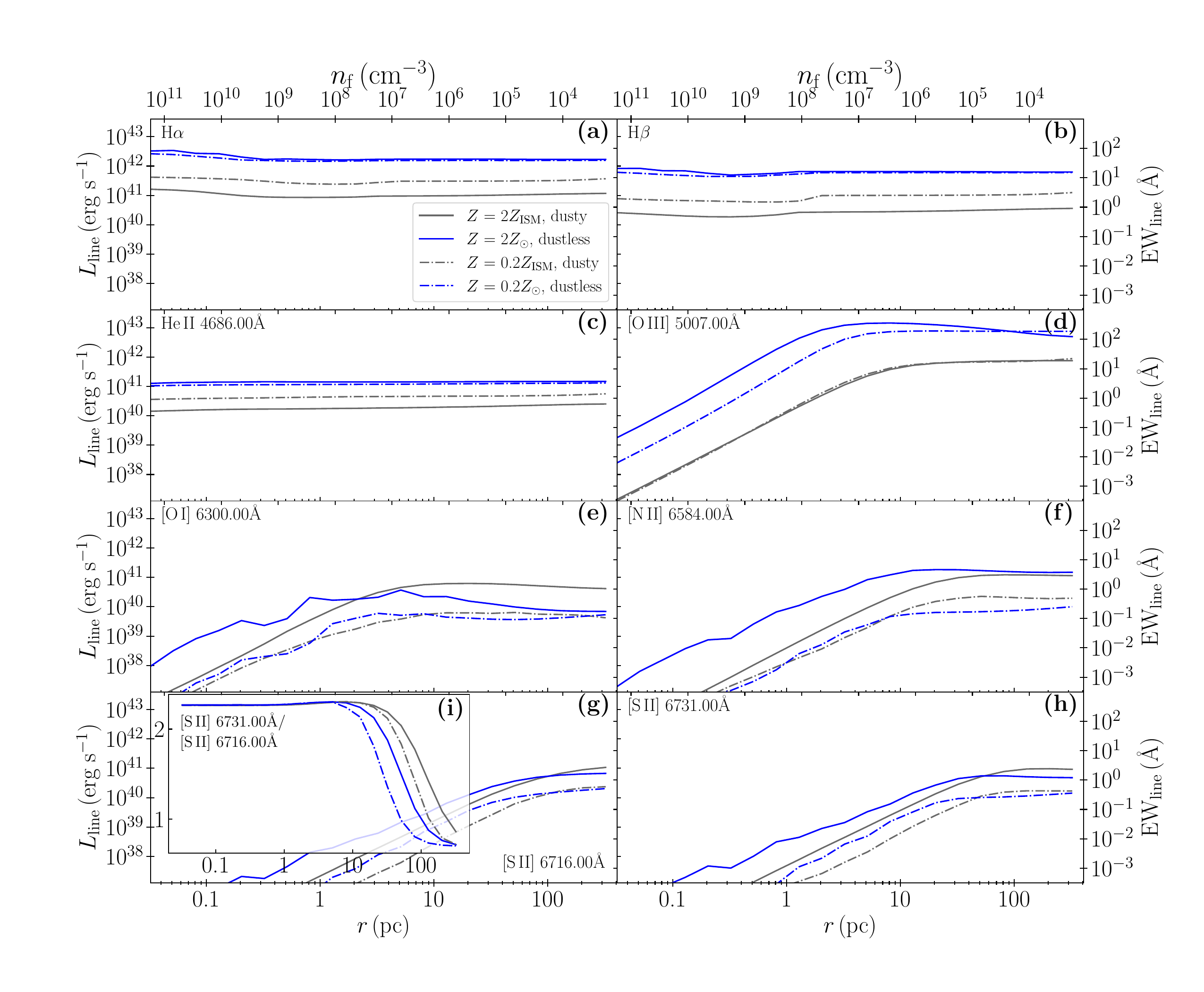}
            \caption{Same as Fig.\ref{fig1} for the optical narrow non CLs and for the recombination lines.  Note the overall more mild dust suppression, by a factor of $\sim 10$ only (no significant depletions to grains), while the lower ionization lines of [N\, II] 6584.00\,\AA\,\,and [S\, II] are not suppressed (at distances where $n_{\mathrm{f}} > n_\mathrm{crit}$), and the lowest ionization line of [O\, I] 6300.00\,\AA\,\,is enhanced. The non-recombination lines peak at larger $r$ compared to the CLs due to their lower $n_{crit}$ values. Subpanel \textbf{(i)} shows that the line ratio [S\, II] 6731\AA/[S\, II] 6716\AA, commonly used as a density indicator, can be used to infer distance.}
            \label{fig2}
        \end{figure*}

        Fig.\ref{fig1}  also presents how the incorporation of dust in the NLR affects the production of optical CLs. Assuming ISM dust, composed of a mixture of amorphous graphite and silicate grains, it depletes metals such as carbon, oxygen, magnesium, silicon, calcium, and iron out of the gas phase and onto the grains \citep{Jenkins2009}. When dust grains are embedded in the photoionized layer, the optical CL emission decreases dramatically, in some cases by three to four orders of magnitude in luminosity and EW. The reduction is produced by two effects. First, the reduction in the thickness of the photoionized layer due to the dominance of the dust in absorbing the ionizing photons (see $\S$\ref{sec:photolay}). Second, the above mentioned depletion of some of the elements from the gas phase. The noble gasses, specifically neon and argon, are not expected to be depleted. Similarly, sulfur does not happen to be part of the grains mineral composition, as a result the strength of the [Ne\,V] 3425.88\,\AA, [Ar\,X] 5533.25\,\AA, and [S\,XII] 7611.00\,\AA\ lines is reduced by only a factor of $\sim 10$, which reflect the reduction in the column of the photoionized layer. Calcium and iron are depleted into grain by up to $\sim 99\%$, leading to a further reduction in strength by an additional factor of $\sim 100$. The presence of grains also lowers the temperature of the highest ionization gas, which leads to further reduction in the line strength, leading to the overall drop by a factor of $10^3-10^4$ in the calcium and iron CLs. The EW of most of the optical CLs in dusty gas is reduced to $\mathrm{EW}\leq 0.01\,$\AA\, which makes them practically undetectable, especially in large detection limited surveys. Only the [Ne\,V] 3425.88\,\AA, is expected to remain detectable with $\mathrm{EW}\sim 1\,$\AA. This result implies that most CL emission observed originates from dustless gas. How gas can potentially become dustless is discussed in $\S$\ref{sec:discussion}. 

        Finally, when comparing the high ($2Z_{\odot}$) to low ($0.2Z_{\odot}$) metallicity calculations, we see that generally a lower metallicity tends to weaken line emission, but the effect is often small. The only exception is with the
        [Ne\,V] 3425.88\,\AA\,\,and [Fe\,VII] 6087.00\AA, the two strongest CL, where the peak emission of dustless gas is suppressed by a factor of $\sim 10$ in low $Z$ gas at $r\sim 1$~pc. The metallicity effect is reduced to a factor of $\sim 2-3$ in the [Fe\,VI] 5176.04, 5335.18\,\AA, and [Fe\,VII] 4839.37\AA\,\,lines in dustless gas.
        In dusty low $Z$ gas some of the line emission is slightly enhanced
        compared to high $Z$ gas due to the associated reduction in the DTG ratio and reduced dust absorption, but the effect is small ($<2$) and is not relevant since most lines are too weak to be detectable (EW $<0.1$\AA).
        
    \subsection{The non-coronal optical lines}\label{sec:NLresults}

        Figure \ref{fig2} presents the predicted luminosity and EW for the lower ionization optical lines, specifically H$\alpha$, H$\beta$, [O~I]~6300\AA, [N\,~II]~6584\AA, 
        [S\,~II]~6716, 6731\AA, and [O\, III] 5007\AA, which are used in the Baldwin-Phillips-Terlevich (BPT; \citealp{1981PASP...93....5B}) diagrams, and also the He\, II 4686\AA\ line. 
        All lines have a maximal EW $>1$\AA, as expected for these most commonly observed narrow lines. By far the strongest line is [O\, III] 5007\AA, which can reach EW $>100$\AA, as indeed observed in some objects \citep[e.g.,][]{BG92}. 
        
        The forbidden lines show the same radial dependence as the optical CLs, with a rise in strength with $r$ followed by flattening. However, the flattening typically occurs at 
        $r\sim 10$~pc or larger, as expected since the critical densities of these lines are significantly lower than those of the optical CLs. The recombination lines, H$\alpha$, H$\beta$ and He\,~II 4686\AA, show little variation as a function of $r$, as expected since all are from permitted transitions and collisional de-exitations do not play a role. 

        Dust suppresses the line strengths by about a factor of $\sim 10$, as found for some of the CLs, and reflects the reduction in the line emitting volume by the dust absorption.
        None of the metals in the non CLs are heavily depleted into the dust grains, and we therefore do not get
        the dramatic line suppression in dusty gas found for some of the CLs. 
        
        Metallicity has no effect on the recombination lines in dustless gas, as these lines just count the number of ionizing photons absorbed by H and He. In dusty gas the dust opacity can dominate, so the reduction in the DTG ratio at low $Z$ enhances the gas absorption which increases the recombination lines strength. The effect of metallicity on the forbidden lines is minimal for [O\, III] 5007\AA, as it remains the main gas coolant, but is significant for the other forbidden lines which are not the main coolants. The effect is most dramatic (factor of $\sim 30$) for the [N\, II] 6584\AA\ line, which reflects its
        steeper than linear scaling with $Z$, and explains why its ratio to H$\alpha$ is a commonly used metallicity indicator \citep[e.g.][]{Kewley2013}. 

        Overall, the traditional NLs do not show the dramatic effect of dust on their strengths, mostly because
        of the lack of a strong depletion of their elements to grains.

        \begin{figure*}
            \centering
            \includegraphics[scale = 0.5]{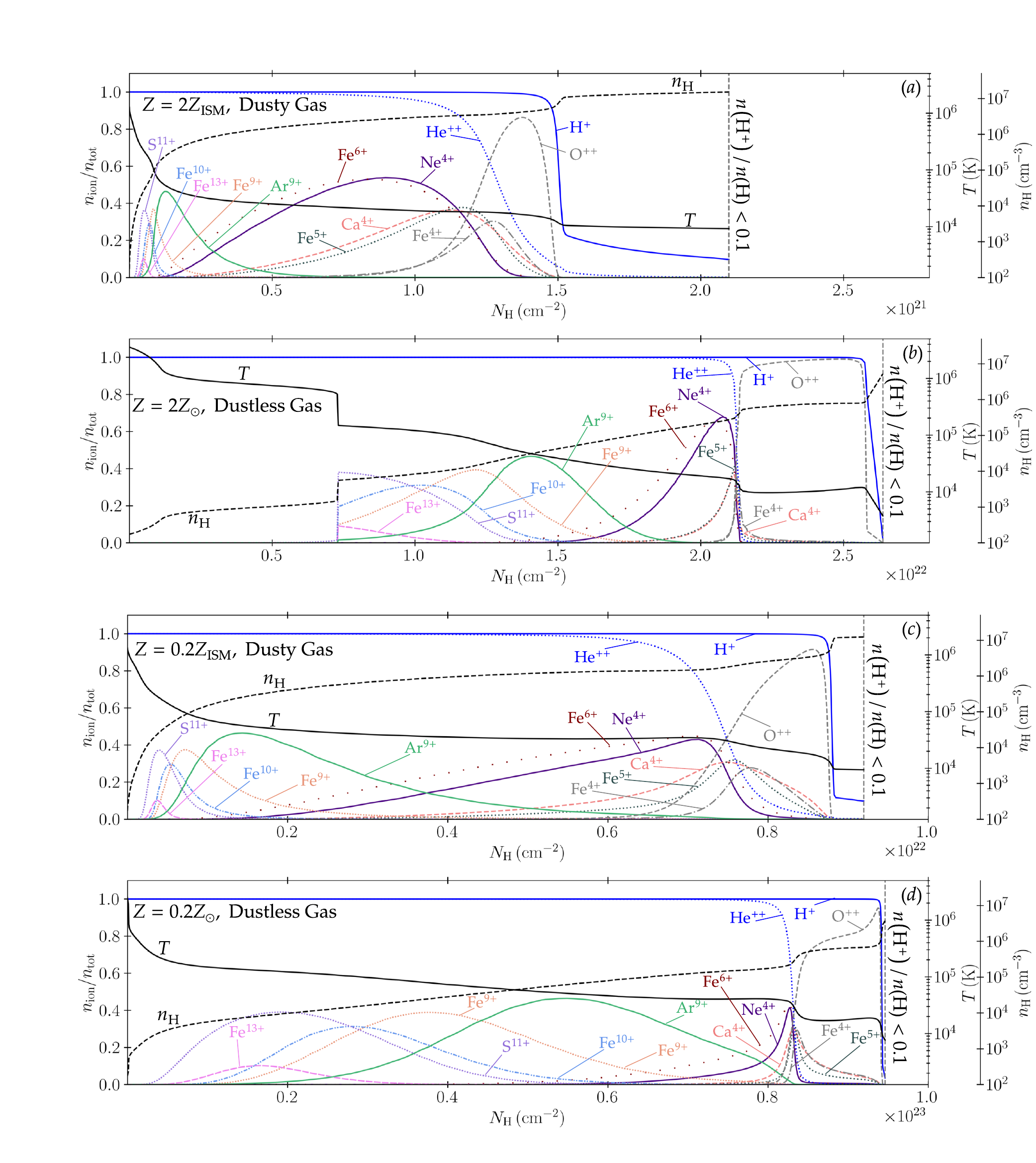}
            \caption{The ionization structure of an RPC slab for dusty and dustless gas at high and low $Z$ values. The slab is placed at $r = 3.25$\,pc and with an ionizing continuum slope of $\alpha_{\mathrm{ion}} = -1.6$. The panels show the various ions abundances relative to the total elemental abundance for each of the CL emitting ions, as a function of $N_{\rm H}$ into the slab, up to the ionization front (location is model dependent).  The panels also show $T$ and $n_{\rm H}$ as a function of  $N_{\rm H}$ (two y-axes on the right). Dust reduces the thickness of the photoionized layer by a factor of $\sim 10$ compared to dustless gas, for both low and high $Z$. High $Z$ reduces the column by a factor of $\sim 4$ compared to low $Z$, for both dusty and dustless gas. The emitting volume of the CL in high $Z$ dusty gas is strongly reduced  compared to the other three models. The largest volume and strongest CLs are expected for dustless gas with a high $Z$.}
            \label{fig3}
        \end{figure*}

        Panel \textbf{(i)} of Figure\,\ref{fig2} shows the [S\, II] 6731\,\AA/[S\, II] 6716\,\AA\ line ratio, which is commonly used as a gas density indicator based on uniform density photoionization models. Since the density is neither a free parameter nor uniform when the RPC effect is taken into account, the ratio  can be used to measure or constrain the distance of the gas from the ionizing source. As panel \textbf{(i)} shows, this line ratio is a useful distance indicator at $r\sim 5-100$~pc,
        where the density is intermediate between the critical densities of the two lines. As the panel shows, the range of $Z$ and dust content explored here lead to a factor of $\sim 2$ uncertainty in the derived $r$ for a given line ratio. Below we explore the dependence of the relative strength of the non CL to the recombination lines, as presented in the BPT diagrams,  on the various model parameters.

    \subsection{The structure of the photoionized layer}\label{sec:photolay}

        Figure \ref{fig3} presents a detailed view of the inner structure of an RPC photoionized slab for the four cases explored above, dusty and dustless gas and each with $Z=2Z_{\odot}$ or $Z=0.2Z_{\odot}$, for a slab of gas placed at $r=3.25$~pc ($10^{19}$~cm) from the ionizing source. 
        Each panel shows the photoionized gas temperature, density, and ionization structure
        as a function of the total hydrogen column density ($N_{\rm H}$) into the slab. The plots also include the relative
        abundance of the specific ions which produce the CLs, recombination lines, and the 
        [O III] line presented above. These plots help understand the effects of the dust and the metallicity on the various lines strength. 
        
        Fig.\ref{fig3} panels \textbf{(a)} and \textbf{(b)} compare the structure of dusty and dustless gas slabs at $Z=2Z_{\odot}$.
        The inclusion of dust decreases the column density of the H ionization front (defined by the sharp drop in the fraction of H$^{+}$) by a factor of $\sim 10$, from $N_{\mathrm{H}}= 2.6\times 10^{22}\,\mathrm{cm}^{-2}$ in dustless gas to $N_{\mathrm{H}}= 1.5\times 10^{21}\,\mathrm{cm}^{-2}$ in dusty gas. A similar factor of $\sim 10$ drop in the photoionized column is present in the $Z=0.2Z_{\odot}$ case, where $N_{\mathrm{H}}$ drops from 
        $9.5\times 10^{22}\,\mathrm{cm}^{-2}$ in dustless gas (panel  \textbf{(c)}) to
        $9\times 10^{21}\,\mathrm{cm}^{-2}$ in dusty gas (panel  \textbf{(d)}). 
        Dust therefore shrinks the thickness of the photoionized layer which affects all line formation, including the optical CLs.
        
        The factor $\sim 10$ reduction in the ionization front location in dusty gas can be understood from the following simple scaling relations.
        Dust competes with gas phase elements for absorbing ionizing photons. The characteristic absorption cross section of dust for ionizing photons is $\sigma_{\mathrm{dust}}\approx 10^{-21}\,$cm$^{2}$, while the absorption cross section of the gas can be approximated by $\sigma_{\mathrm{gas}}\approx 10^{-18}n(\mathrm{H}^{0})/n(\mathrm{H})\,\mathrm{cm}^{2}$, where $n(\mathrm{H}) = n(\mathrm{H}^{0}) + n(\mathrm{H}^{+})$ \citep{1993ApJ...402..441L}. In the photoionized region, $n(\mathrm{H}^{+})/n(\mathrm{H}^{0})\approx 10^{5}U$ \citep{1993ApJ...402..441L}. So, the ratio of dust to gas opacity is:
        \begin{equation}
            \frac{\tau_{\mathrm{dust}}}{\tau_{\mathrm{gas}}} = \frac{\sigma_{\mathrm{dust}}}{\sigma_{\mathrm{gas}}} \approx 100U.
        \end{equation}
        Therefore, dust opacity dominates over gas opacity when $U\gtrsim10^{-2}$ (see \citealp{1993ApJ...404L..51N} for a more rigorous calculation). In a dusty RPC slab, this is always true as $U$ has the universal value of $\sim$0.01 at the ionization front \citep{2014MNRAS.438..901S}, the location where the radiation pressure is nearly fully converted to gas pressure. Since the dust UV opacity is $\sim$10 times larger than the gas opacity, $\sim 90$\% of the ionizing photons are absorbed by dust, and only $\sim 10$\% are absorbed in the gas phase leading to line emission. This effect leads to a factor of 
        $\sim 10$ reduction in the line emission. 
        
        As discussed above in $\S$\ref{sec:CLresults} an additional significant reduction in some of the line emission is produced for elements which are heavily depleted from the gas phase to the grains, in particular the calcium and iron lines. In addition the dusty photoionized slab is generally colder due to the grain cooling of the gas, 
        and as a result denser than a dustless slab. These changes affect the ionization structure within the slab and produce more minor changes in the various line strengths in dusty versus dustless gas. 

        The effect of metallicity changes can be derived by comparing the dusty gas simulation with $Z=2Z_{\odot}$ and $Z=0.2Z_{\odot}$
        in panels \textbf{(a)} and \textbf{(c)}. The column density at the ionization front increases
        from  $N_{\mathrm{H}}= 1.5\times 10^{21}\,\mathrm{cm}^{-2}$ to 
        $9\times 10^{21}\,\mathrm{cm}^{-2}$. This factor of 6 increase in the column occurs since the DTG ratio
        scales linearly with $Z$ \citep{2013A&A...557A..95R, 2015MNRAS.449.3274F, 2019A&A...623A...5D, Galliano2021}, so the drop by a factor of 10 in the dust opacity makes it comparable with the gas opacity, leading a total opacity drop by a factor of $\sim 6$. 
        
        The effect of metallicity in dustless gas can be derived by comparing panels \textbf{(b)} and \textbf{(d)}. The ionized column increases from 
        $N_{\mathrm{H}}= 2.1\times 10^{22}\,\mathrm{cm}^{-2}$ to 
        $9.5\times 10^{22}\,\mathrm{cm}^{-2}$. The non linear increase of $N_{\mathrm{H}}$ with $1/Z$ reflects the changes in the temperature and density structure of the slab, which affect the gas opacity and the relative contribution of H and He which increases when Z decreases. 

        We note in passing that since the dust opacity in the optical regime is $\sim 1/10$ of the dust UV opacity \citep{1993ApJ...402..441L}, and the slab UV opacity is $\sim 1$, the optical lines produced in the slab are not significantly absorbed by the dust on their way out. The dominant effect of the dust is the absorption of the incident ionizing UV radiation. Clearly, the detailed photoionization modeling presented above are required in order to derive accurately the effects of dust and metallicity on the various emission lines.

        \begin{figure*}
            \centering
            \includegraphics[scale = 0.5]{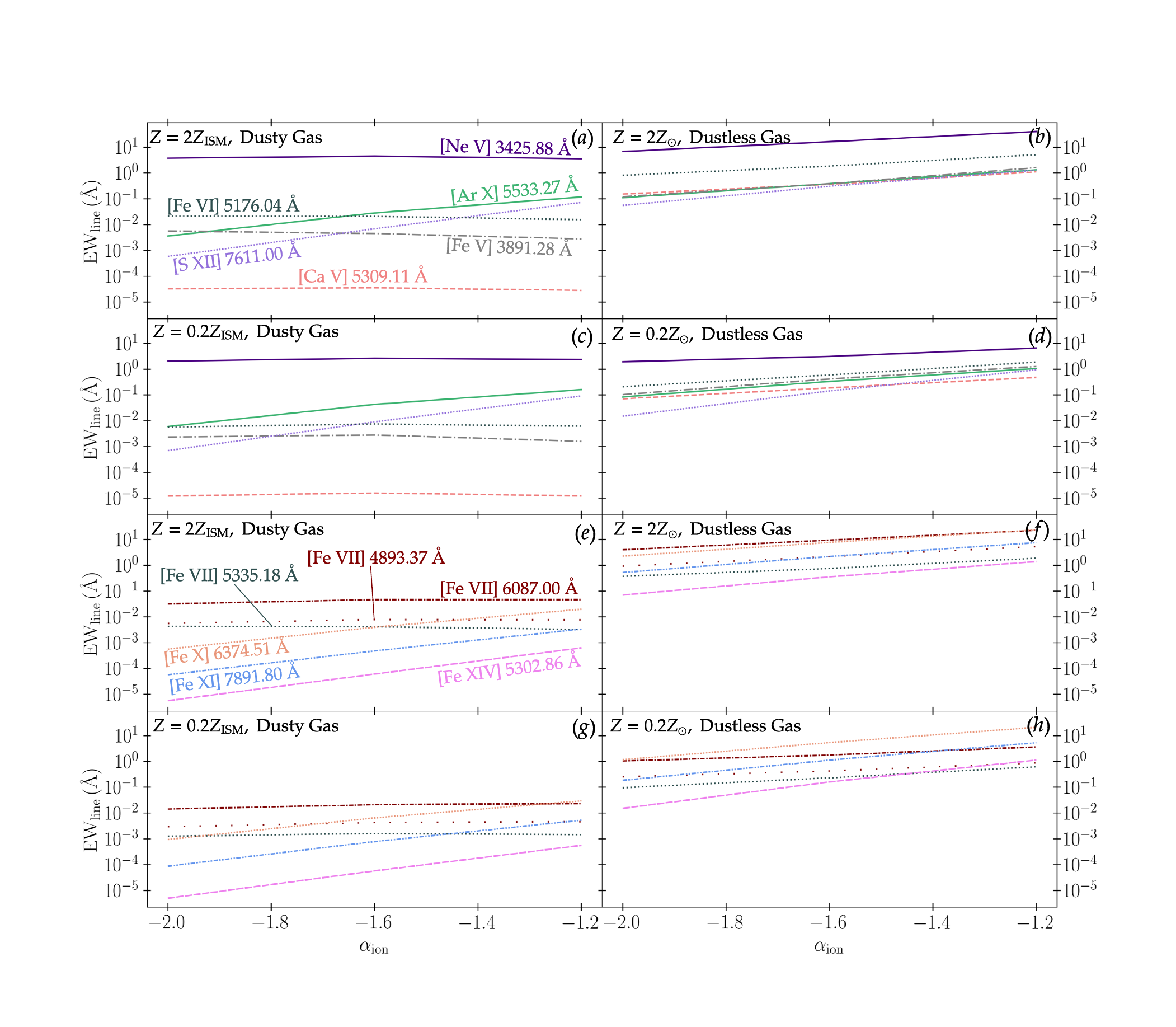}
            \caption{The EW of the optical CLs as a function of $\alpha_{\mathrm{ion}}$. The models assume dusty (left panels) and dustless (right panels) gas with low and high $Z$, placed at $r = 3.25$\,pc. In dusty gas, the relatively lower ionization level CLs of [Ne\,V], [Ca\,V] and [Fe\,V], [Fe\,VI], and [Fe\,VI], do not increase, or even decrease, with increasing $\alpha_{\mathrm{ion}}$. While the higher ionization CL from [Ar\,X], [S\,XII], and [Fe\,X], [Fe\,XI], and [Fe\,XIV], increase significantly with $\alpha_{\mathrm{ion}}$. However, only the [Ne\,V] 3425.88\AA\,\,line is clearly detectable (EW$>1$\,\AA) in dusty gas. In dustless gas essentially all lines are detectable  (EW$>0.1$\,\AA), and their EW increases with $\alpha_{\mathrm{ion}}$ (with different slopes). The strongest CLs are expected from dustless gas illuminated by a hard ionizing continuum.}
            \label{fig4}
        \end{figure*}
        \begin{figure*}
            \centering
            \includegraphics[scale = 0.5]{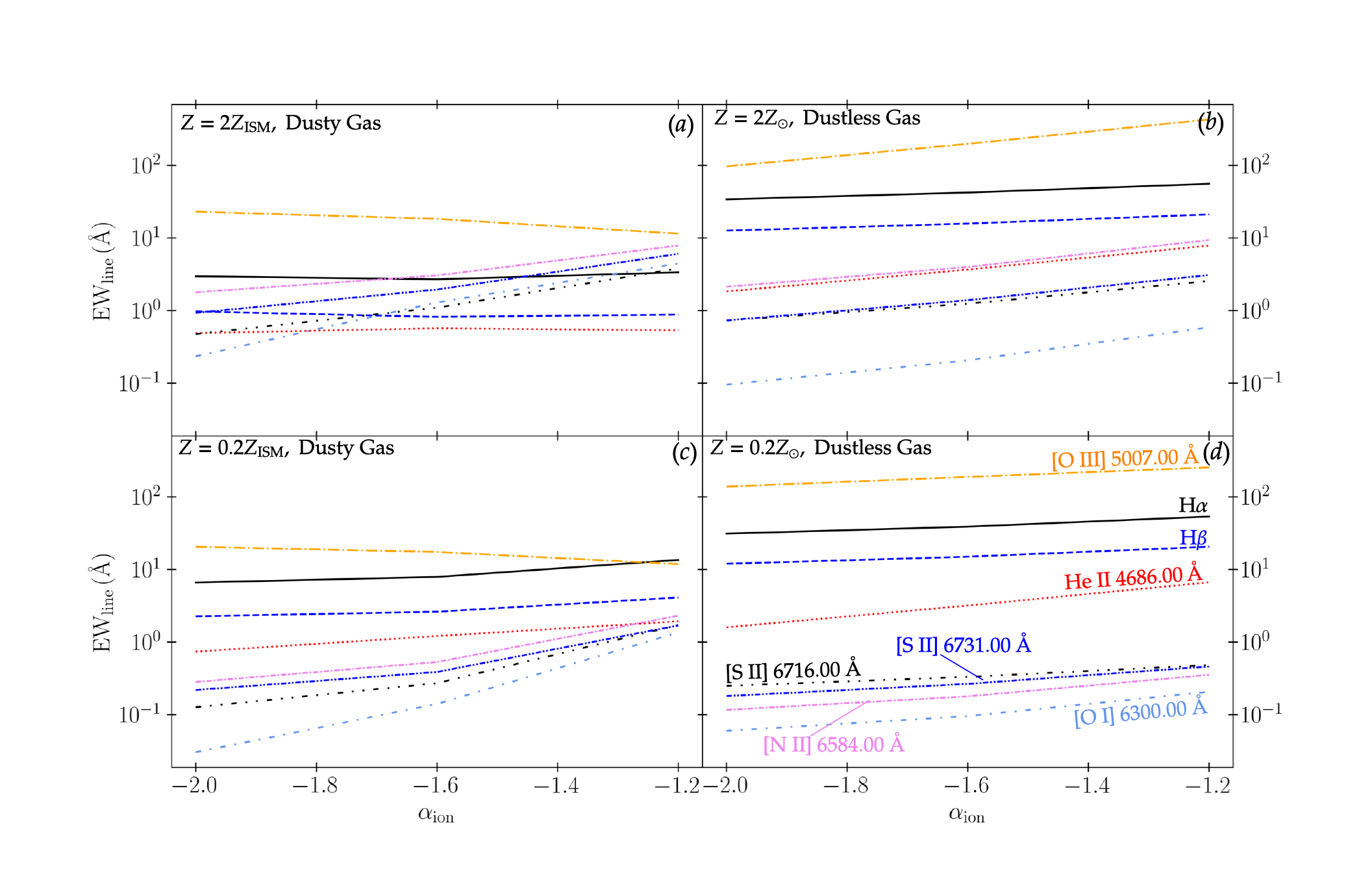}
            \caption{Same as Fig.\ref{fig4}, for the optical non CLs and for the optical recombination line. The RPC slab is located at $r = 32.5$\,pc, where the densities for all these lines are below their critical densities. The EW of the strongest NL, [O\, III] 5007\,\AA, and also the hydrogen and helium recombination lines, decrease with increasing $\alpha_{\mathrm{ion}}$ in dusty gas, and increases in dustless gas. Exceptionally large EW of the [O\, III] 5007\,\AA\,\,line may serve as a marker for strong CLs. All the other optical non CLs, increase in EW with increasing $\alpha_{\mathrm{ion}}$, and the effect is stronger in dusty gas.}
            \label{fig5}
        \end{figure*}

    \subsection{The effect of the SED on the Coronal and non Coronal lines strength}\label{sec:SED}
    
        Figure \ref{fig4} presents the dependence of the strength of the 12 optical CLs on the hardness of the ionizing continuum for a slab located at $r= 3\,\mathrm{pc}$. The figure shows the EW as a function 
        of $\alpha_{\mathrm{ion}}$ in the range of $-2.0$ to $-1.2$, which covers the observed range of spectral slopes in AGN \citep[e.g.,][]{Steffen2006,Just2007}. This slope extends from 13.6~eV
        to 1~keV, and the observed range of slopes corresponds to a large range in the X-ray/UV luminosity ratio.
        Specifically, the flattest slope observed corresponds to a ratio which is a factor $\sim 30$ larger than in the steepest slope objects. The slope is tightly correlated with $L_{\rm bol}$ and the flattest, hardest slope is found in lowest luminosity AGN, and may be related to the lowest BH mass AGN. The eight panels
        in Fig.\ref{fig4} present the EW versus $\alpha_{\mathrm{ion}}$ for the [Ne\,V], [Fe\,V], [Ca\,V], [Fe\,VI], [Ar\,X] and [S\,XII] lines (upper four panels), and for the [Fe\,VII], [Fe\,X], [Fe\,XI] and [Fe\,XIV] lines (lower four panels). For each set of lines, the panels cover the four cases of dusty and dustless gas, each for $Z=2Z_{\odot}$ and $Z=0.2Z_{\odot}$.
        
        The high ionization state of the ions which produce CL suggests that the strength of these lines will be highly dependent on the hardness of the ionizing continuum. Indeed, a strong dependence exists for the 
        [Ar\,X]~ 5533.27\AA\ and [S\,XII]~7611.00\AA\ lines in dusty gas at low and high $Z$, where the EW increases by a factor of $\sim 50-100$ when the SED changes from soft $\alpha_{\mathrm{ion}}=-2.0$ to hard $\alpha_{\mathrm{ion}}=-1.2$ (panels \textbf{(a, c)}). A similar increase is present for
        the  [Fe\,X] 6347.51\AA, [Fe\,XI] 7891.80\AA, and [Fe\,XIV] 5302.86\AA\ lines in dusty gas (panels \textbf{(e, g)}).
        However, the [Ne\,V]~3425.88\AA, [Ca\,V] 5309.11\AA\ and the [Fe\,VII]~4893.37, 5335.18, 6087.00~\AA\ lines
        are not affected at all by the changes in $\alpha_{\mathrm{ion}}$, while the lower IP lines [Fe\,V]~ 3891.28\AA\ and 
        [Fe\,VI] 5176.04\AA\ lines actually get weaker when considering the harder SED. 

        In dustless gas (right panels of Fig.\ref{fig4}), essentially all lines show a rise in EW with
        $\alpha_{\mathrm{ion}}$, though the rise is of a somewhat smaller amplitude (e.g., a factor of
        $\sim 10$ instead of $\sim 50$ for the [Ar\,X] 5533.27\AA\ line, and $\sim 30$ instead of $\sim 100$
        for the [S XII]~7611.00\AA\ line). However, as described above ($\S$ \ref{sec:CLresults}) the 
        derived EWs of the optical CLs in dusty gas, excluding the [Ne\,V]~3425.88~\AA\ line, 
        are mostly $\ll 0.1$~\AA. Despite the more significant enhancement of some of the CLs in the 
        dusty gas, only the [Ar\,X] 5533.27\AA\ line reaches EW$=0.1$~\AA\ at $\alpha_{\mathrm{ion}}=-1.2$,
        and the other lines remain non detectable even for the hardest observed SED. 

        Thus, the critical factor to have detectable CLs is emission from dustless gas.
        Once the gas is dustless, a harder SED clearly further enhances the strength of the CLs 
        by a factor of $\sim 10$, compared to the softest SED. Already at the intermediate slope of $\alpha_{\mathrm{ion}} = -1.6$, most lines become observable, and the strongest CLs are expected
        for dustless gas illuminated by a hard SED.
        
        Figure \ref{fig5} presents the dependence of the strength of the recombination and non CLs 
        on the hardness of the ionizing SED. The SED effect is generally smaller than found for the CLs.
        The H$\alpha$ and H$\beta$ recombination lines either decrease slightly or increase by $< 2$ with 
        $\alpha_{\mathrm{ion}}$. A larger effect is present for the He~II~4686.00\AA\ recombination line,
        which increases by up to a factor of $\sim 5$ for dustless low $Z$ gas. The [O III]~5007.00\AA, which is 
        the  strongest NL, shows opposite dependencies. In dusty gas it decreases by a factor of $\sim 2$
        with increasing $\alpha_{\mathrm{ion}}$, while in dustless gas it increases by a factor of $2-5$. 
        The lower ionization forbidden lines, [N II], [S II] and [O I], generally increase with $\alpha_{\mathrm{ion}}$, by about a factor of $2-4$ in dustless gas, and by a factor of 
        $10-40$ is dusty gas. The strongest low ionization lines are obtained for a hard continuum in 
        high $Z$ dusty gas, the conditions where the [O\, III]~5007.00\AA\ is weakest (Fig.\ref{fig4}, panel
        {\bf(c)}).

        Overall, the lower ionization forbidden lines, [N II], [S II] and [O I], are expected to be
        weakest in dustless gas, the conditions where the CLs are expected to be strongest. Only the 
        [O III]~5007.00\AA\ line becomes strongest for dustless high $Z$ gas with a hard SED, the same
        conditions where the CLs are strongest. This is not surprising as the $\mathrm{O}^{+2}$ ion is present
        in the $\mathrm{H}^+$ region, in front of the ionization front, the region where all the CL are produce
        (Fig.\ref{fig3}), while the singly ionized and neutral ions which produce the lower ionization 
        lines, are generally present very close or behind the ionization front 
        \cite[e.g.][Figure A1 there]{2014MNRAS.438..901S}.
        Therefore, it is possible that strong [O\, III] 5007\AA\ emission is a necessary (but not sufficient) indicator 
        for significant optical CL emission, as indeed suggested by some observations \citep{Thomas2017}.

        \begin{figure*}
            \centering
            \includegraphics[scale = 0.5]{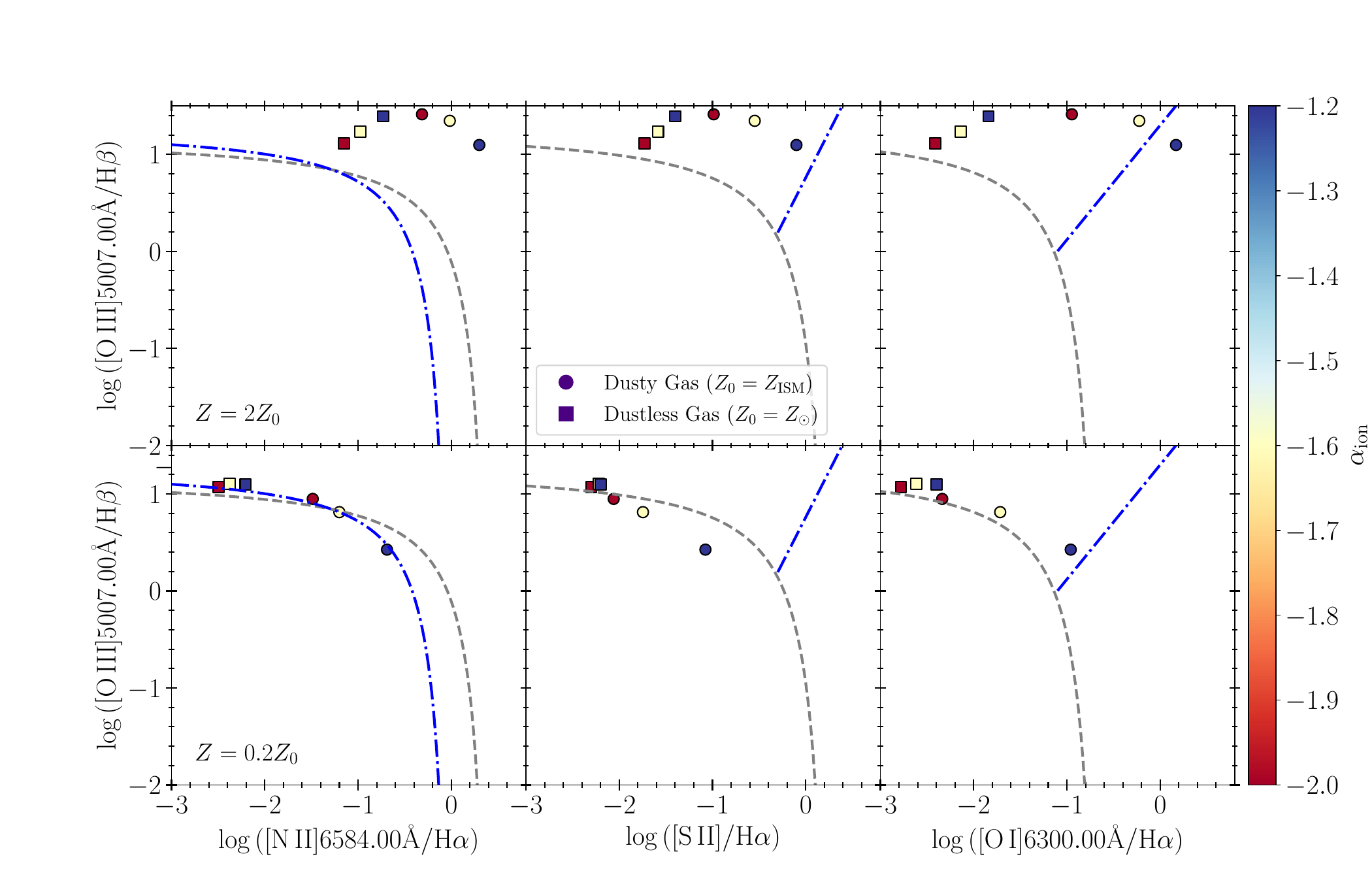}
            \caption{The BPT diagrams for dusty and dustless gas at high (upper panels) and low (lower panels) $Z$, for a RPC slab placed at $r =32.5$\,pc, exposed to different ionizing slopes (right color bar). The dashed black line is the \cite{Kewley2001} theoretically derived separation line of AGN from starburst galaxies. The dashed dotted blue line (left BPT panel only) is the similar empirically derived line of \cite{Kauffmann2003}. The 
            RPC results for low $Z$ gas happens to coincide with the empirically Kauffmann line, suggesting pure AGN 
            excitation can extend up to the starburst region for low $Z$ AGN.  Higher $Z$ AGN span the more populated AGN area in these BPT plots. Strong CL AGN are expected to populate the left sides of the BPT plots, where the dustless models reside.}
            \label{fig6}
        \end{figure*}

    \subsection{The BPT locations of the coronal line emitting gas}    

        Figure \ref{fig6} presents the BPT diagrams for high $Z$ (upper panels) and low $Z$ (lower panels) 
        gas placed at $r=32.5\,$pc. Each panel shows the results for dusty and dustless gas,
        each with $\alpha_{\mathrm{ion}}=-1.2, -1.6$ and $-2.0$. This value of $r$ is chosen to ensure that $n_{\mathrm{f}} < n_{\mathrm{crit}}$ so the line ratios correspond to the regions which dominate the line emission. The high $Z$  model results generally fall in the AGN region in all three BPT line ratio panels. The only exception is in the $\alpha_{\mathrm{ion}} = -1.2$ case for the [O\, I] 6300\,\AA/H$\alpha$ ratio, which resides in the LINER region of the diagram. The low $Z$ calculations show that dusty RPC gas with hard ionizing slopes falls in the starburst region of the [S\, II]/H$\alpha$ and [N\, II] 6584\,\AA/H$\alpha$ BPT panels. However, the AGN/starburst boundary (dashed gray lines) is based on the
        \cite{Kewley2001} photoionizing shocks modeling results, which are different from the Cloudy RPC
        photoionization results used here. So, this apparent mismatch just reflects the models mismatch. The 
        low $Z$ RPC models results happens to coincide with the empirically derived \cite{Kauffmann2003} 
        AGN/starburst boundary in the [N\, II] 6584\,\AA/H$\alpha$ BPT panel (dash dotted blue line).
        The region between the Kewley and the Kauffmann lines is commonly assumed to harbour composite objects,
        where AGN and starbursts contribute similarly. Our results suggest that objects in the composite region
        can be powered by pure AGN emission from low $Z$ gas.
          These results are in good agreement with the theoretical trends found in \cite{2019ApJ...870L...2C}. 

        Where should the CL AGN be preferentially found on the BPT diagrams? The gas clearly needs to be dustless (see $\S$\ref{sec:SED}), otherwise the EW of all CL remains $<0.1$\AA, regardless of $Z$ and $\alpha_{\rm ion}$, and are likely non-detectable (Figs.\ref{fig1},\ref{fig4}), excluding the [Ne V] line). The lack of dust
         implies an offset to the left in all the BPT diagrams. This trend is consistent
        with the results of \cite{2022ApJ...936..140R} that CL emitters preferentially reside in the star forming regime with $-1.5\lesssim\log{(\text{[N\, II]\,\,6584\,\AA/H}\alpha)}\lesssim -0.5$.

        We note in passing that the range of metallicities, SEDs, and dust probed here, do not lead to models which populate 
        the LINER regime in the BPT diagrams. The exceptional strength of the low ionization NLs in LINER 
        galaxies may reflect a very hard ionizing
        continuum, which peaks in the X-ray regime and powers the line emission from the low ionization region.
        We note that the SEDs explored in this study are not applicable to LLAGN, which constitute the dominant population in the local universe \citep{Ho2008}. Additional emission mechanisms can also be present in LINERs such as those from shocked gas (e.g., \citealp{Molina2018}). We consider the possible production of CLs from shocked gas in $\S$\ref{shockCalc}

\section{Discussion} \label{sec:discussion}
        
    \subsection{Which conditions are needed for strong optical CL production?}\label{sec:DustDest}

        The RPC photoionization models presented above indicate that a CLER is in fact present in any photoionized gas slab exposed to even a moderate AGN SED (See Fig.\ref{fig3}), in contrast with constant density photoionization models \citep[see a comparison in][]{2014MNRAS.438..604B}, where one needs to invoke a separate high $U$ gas component to produce CLs.  
        
        Optical CL are not always detectable by current surveys (e.g., SDSS). The free parameters which affect their strength are (1) the gas phase metallicity, (2) the presence of dust, and (3) the hardness of the ionizing continuum, all of which we explored here. These models have shown that the critical parameter in the production of strong CL emission is the presence of dust. We have demonstrated that CLs are mostly non-detectable at the sensitivity limit of current large-scale optical surveys in dusty gas ([Ne V] 3425.88\AA\ is the only exception). The faintness of the lines results from two effects: (1) dust grains absorb $\sim 90$\% of the ionizing photons, and (2) the heavy depletion of $\sim 99$\% for some of the primary elements, like calcium and iron, from the gas phase. These effects reduce the CL strengths in dusty gas by factors of $10-10^3$ in luminosity and EW (see Fig.\ref{fig1}), with the largest reductions seen in the most refractory elements.  The [Ne V] 3425.88\AA\ line remains the strongest line since neon, a noble gas, is not depleted. The hardness of the ionizing SED can also affect the CL stength, but has a secondary effect. Using our hardest SED, the CL strengths can be up to a factor of $\sim 10$ larger than that produced from the softest SED model (Fig.\ref{fig4}). While a harder SED can enhance the CL strengths, our models indicate that CLs will nonetheless be undetectable in current surveys unless the gas is dustless.  Thus, this work indicates that the strongest CLs are expected in AGN with a hard ionizing SED and dustless gas. 

        The RPC results show that indeed the [Ne\,V] 3425\,\AA\ and the [Fe\,VII] 6087\,\AA\ lines are expected to be the strongest
        CLs, which can reach EW$\sim 10$\AA, and $F(\text{[Fe\,VII] 6087\,\AA}) / F(\text{H}\alpha)>0.25$ and $F(\text{[Ne\,V] 3425\,\AA}) / F(\text{H}\alpha)>0.2$ (Figs.~\ref{fig4}, \ref{fig5}), as observed in CLiF AGN \citep{2015MNRAS.451L..11R, 2015MNRAS.448.2900R, 2016ApJ...824...34G, CC2021}.

    \subsection{The origin of dust-free gas}
    
        Although evidence for dusty gas in the central regions of AGNs has been found \citep[e.g.,][]{1993ApJ...402..441L, 1993ApJ...404L..51N, 2003ApJ...587..117R, 2023ApJ...954..175Z}, some have argued that dust-free gas models provide more accurate predictions compared to the dusty models based on the observations of UV and optical high-ionization forbidden emission lines (e.g., \citealp{2003AJ....125.1729N, 2006A&A...447..863N}). For optical CLs specifically, the absence of dust for detectable emission has been suggested theoretically (e.g., \citealp{1995ApJ...450..628P, 1997ApJS..110..287F}) and found observationally (e.g., \citealp{2023ApJ...945..127N}). Gas in AGN becomes dustless at distances smaller than the dust sublimation radius,
        which corresponds to the size of the BLR \citep{1993ApJ...402..441L, 2018MNRAS.474.1970B}. The gas density at the sublimation radius can be estimated as follows (see also \citealp{2018MNRAS.474.1970B}). The flux required to sublimate a dust grain (assuming a strictly spherical grain emitting as a blackbody) is $F_{\mathrm{sub}} = 4\sigma T_{\mathrm{sub}}^{4}$ where $\sigma=5.670\times 10^{-5}\mathrm{erg}\,\mathrm{s}^{-1}\,\mathrm{cm}^{-2}\,\mathrm{K}^{-4}$ is the Stefan-Boltzmann constant and $T_{\mathrm{sub}}$ is the grain sublimation temperature. In dense gas $T_{\mathrm{sub}}\approx 2000\,$K (e.g., \citealp{2018MNRAS.474.1970B}), the radiation pressure at the sublimation radius will be $P_{\mathrm{rad}} = F_{\mathrm{sub}} / c \approx 3.630\times 10^{9}\,$erg\,cm$^{-3}$. At the ionization front, $P_{\mathrm{rad}} = P_{\mathrm{gas}} = 2.3n_{\mathrm{H}}k_{\mathrm{B}}T_{\mathrm{sub}}$ (eq.\ref{eqU}) which gives $n_{\mathrm{H}}=2\times$10$^{11}$cm\,$^{-3}$. Therefore, most CLs are collisionally suppressed at such a small radius (i.e., the BLR radius $<0.1$~pc, Fig.\ref{fig1}), and would be too weak (EW$\sim 0.1$~\AA) to be detectable if they originate at the dust sublimation radius \citep{Glidden_2016}. Indeed, optical CL production is observed out to hundreds of parsecs (e.g., \citealp{2010MNRAS.405.1315M, 2018ApJ...858...48M, 2021ApJ...920...62N, 2023ApJ...945..127N}), so dust sublimation cannot produce the required dustless gas at these radii. Other dust destruction mechanisms are required to explain the prominent extended CL emission. 

        Thermal sputtering, grain destruction via collisions with surrounding gas particles, becomes efficient
        in gas with $T\gtrsim 10^6$\,K \citep[Figure 7 in][]{Draine1979},
        and has also been suggested as a means of CL production in AGN (e.g., \citealp{1995ApJ...450..628P}).        
        Diffuse gas in the AGN host galaxy ISM can reach $\gtrsim 10^{6}\,$K, and cause efficient sputtering, at distances as large as $\sim$ 1\,kpc from the center \citep{2006ApJ...645.1188W} . In dusty RPC gas, the required temperature needed for efficient sputtering is achieved only at a very small optical depth 
        $\tau\ll 0.1$ ($N_{\mathrm{H}}\ll 10^{20}\,$cm$^{-2}$), while the high ionization ions with IP$ \gtrsim 100\,$eV are present at $\tau\sim 0.3-1.5$ ($N_{\mathrm{H}}\sim 10^{20-21}\,$cm$^{-2}$) where $T\lesssim 10^5$\,K (Fig.\ref{fig3}). Therefore thermal sputtering may not be effective in destroying grains in the RPC slab region where CLs are being produced. 

        Strong shocks have been shown to destroy dust grains in the ISM (e.g., \citealp{1998ApJ...501..643D, 2004ASPC..309..347J, 2015ApJ...803....7S}; Matzko et al.\,submitted). This could be the case in AGNs, as CLs have been shown to align with radio jets in some sources (e.g., \citealp{2005MNRAS.364L..28P, 2023arXiv230713263N}) but not all (e.g., \citealp{2010MNRAS.405.1315M}). AGN outflows have also been shown to have a low reddening value and thus may be dust free (e.g., \citealp{2003ARA&A..41..117C, 2014MNRAS.445.3011S}). \cite{2021ApJ...911...70B} have shown that strong CLs are detected in dwarf galaxies with outflows. The CL ratios are consistent with shock models
        \citep[e.g.][but see discussion below]{Wilson1999, 2006ApJ...653.1098R, 2011A&A...533A..63E, 2021ApJ...922..155M}, possibly suggesting grain destruction through shocks in the outflowing gas. Large-scale statistical studies of CL emitters have also recently revealed a statistically significant increase in outflows in CL emitters in AGNs compared to a matched control sample (Matzko et al.\,\,submitted; Doan et al.\,\,submitted). Thus, we suggest that shocks may be effective in destroying the grains in gas which is subsequently photoionized and produce CLs.
        
        Finally, dust grains can be prevented from forming to begin with. Grain growth in the ISM is tied to the host galaxy metallicity (e.g., \citealp{1980ApJ...239..193D, 2012arXiv1202.2932I, 2011ApJ...735...44Y, 2013EP&S...65..213A}) and therefore low metallicity sources have less dust grains and potentially stronger optical CL emission. In fact, a growing number of optical CLs have been detected in the dwarf-galaxy population \citep{2012MNRAS.427.1229I, 2021ApJ...911...70B, 2021ApJ...912L...2C, 2021MNRAS.508.2556I, 2021ApJ...910....5M, 2023ApJS..265...21R, 2023arXiv230502189H, 2023ApJ...946L..38R} which reside in a metallicity regime where the DTG ratio is potentially non-linear in $Z$ ($Z\sim$0.2$Z_{\odot}$; e.g., \citealp{2013A&A...557A..95R, 2015MNRAS.449.3274F, 2019A&A...623A...5D}, Galliano2021), and a drop in the DTM, that is a lower depletion into grains which will lead to an enhancement of optical CL emission. 
        
        Prominent He\, II 4684\,\AA\,\,emission has also been observed in metal poor galaxies with line flux ratios relative to the hydrogen recombination lines too high to be explained by photoionization models with even the most extremely metal-poor stars (e.g., \citealp{2018ApJ...859..164B, 2019MNRAS.490..978P, 2021ApJ...908...68O, 2021A&A...656A.127S}) suggesting an additional source of ionization. This raises the possibility that CL emission can uncover accretion activity in the low-metallicity galaxies where accreting IMBHs reside, and where traditional optical narrow line diagnostics fail \citep{2019ApJ...870L...2C}. We also note that searches for optical CLs with JWST at high redshift, where rest-frame optical and UV spectra are available, have revealed weak or non-detections \citep{2023arXiv230200012K, 2023arXiv230512492M, 2023arXiv230206647U}. While there is a trend of decreasing host galaxy metallicity with redshift (e.g., \citealp{2023ApJ...957...39L}), the nuclear regions of AGN at high $z$ could still have high metallicity and thus dust content (e.g., \citealp{1999ARA&A..37..487H}). The lack of optical CLs in high $z$ AGN may then point to rapid dust formation in the early universe, as ALMA (e.g., \citealp{2015Natur.519..327W, 2020A&A...643A...2B, 2020A&A...643A...5D}) and early JWST observations suggest (e.g, \citealp{2023Natur.621..267W}).  
        
       \subsection{Can shock excitation contribute to the CL emission?}\label{shockCalc}  

        As described above, dust destruction by shocks can produce the low DTG and DTM ratio (i.e., low metal depletion to dust grains) which characterizes CL emitting gas. However, shocks also power line emission \citep[e.g.,][]{Allen08}. Can shocks, rather than AGN photoionization drive the observed CL emission \citep[e.g.,][]{Wilson1999}?  The standard approach to distinguish between shocks and AGN photoionization is based on various line ratios, such as the BPT diagrams
        \citep[e.g.][]{Kewley2013}. However, can shocks reproduce the observed line luminosities? Reproducing the observed line luminosities in AGN is challenging with pure shock models, as we discuss below.

        The shock line emission per unit area is set by the gas density $n$ and the shock speed $v_s$. Based on the detailed photoionizing shock modeling of \cite{Allen08}, strong [Ne V] 3425.88\AA\ emission is produced for $v_s=1000$~km~s$^{-1}$ in a model which includes both the shock and the precursor emission (i.e., a photoionizing shock). This model gives [Ne V]/H$\alpha =0.325$, which is as high as obtained above for photoionized gas (see Fig.\ref{fig1} and Fig.\ref{fig2}). The derived [Ne V] 3425.88\AA\ line emission per unit area, $F$ (in erg~cm$^{-2}$~s$^{-1}$) is $\log (F/n)=-2.62$ \citep[see Table 8 in][]{Allen08}. The characteristic FWHM of the [Ne V] line, as observed in large optical surveys such as the CLASS survey \citep{2022ApJ...936..140R} is $\sim 300-600$~km~s$^{-1}$ (although larger values have been reported; see Doan et al., submitted), for which \cite{Allen08} gives $\log (F/n)=-4.80$ to $-3.57$. At $v_s=300-1000$~km~s$^{-1}$ the shock models produce [Ne V]3425.88\AA\ / H$\alpha = 0.031 -  0.11$, which spans part of the observed range of [Ne V]/H$\alpha \simeq 0.07-0.85$ \citep[][Doan et al. submitted]{2022ApJ...936..140R}. However, the shock emission line luminosity is likely too low, as explained below. 
        
        We note in passing that the relative strengths of iron CLs can vary significantly between shock models (e.g., \citealp{Wilson1999}). Additionally, since neon is a noble gas and therefore not subject to dust depletion, we choose to focus our calculations on the [Ne\,V] CL, which is also the strongest optical CL.

        Here we derive the total flux in mass, momentum, and energy through shocks required to produce the observed [Ne V]~3425.88\AA\ 
        line luminosity. The mass flux $\dot{M_s}$ passing through shocks, with a total shock front area of $A$, is
        \begin{equation}\label{Mdot}
               \dot{M_s}=A\times v_s\times n \ .         
        \end{equation}
      To produce a line luminosity $L_{\rm line}$ requires a shock area of $A=L_{\rm line}/F$, and thus
        \begin{equation}\label{Mdot1}
               \dot{M_s}=\frac{L_{\rm line}}{F/n}\times v_s\ ,             
        \end{equation}
      Implementing the \cite{Allen08} results of $\log (F/n)=-4.80,\ -3.57$ and  $-2.62$ for $v_s=300,\ 600$ and $1000$~km~s$^{-1}$,
       using $L_{\rm line}=10^{41}L_{41}$, yields  
        \begin{equation}\label{Mdot2}
               \dot{M_s}= (5014, 591, 110)\times L_{41}\,M_\odot\,{\rm yr}^{-1}\ .            
        \end{equation}
        In contrast, photoionization can produce a [Ne V] 3425.88\AA\ luminosity with 
        $L_{\rm line}\sim (1-10)\times L_{41}$ for an AGN with $L_{\rm bol}=10^{45}$~erg~s$^{-1}$ (Fig.\ref{fig1}), which
        requires a mass accretion rate of only $ \dot{M_{\rm ac}}= 0.18\ M_\odot\ {\rm yr}^{-1}$ 
        (for a 10\% accretion radiative efficiency). The drastically higher mass flux through shocks, $\dot{M_s}\sim (10^3-10^4) \times \dot{M_{\rm ac}}$, is inevitable as the energy released per unit mass is $\sim (v_s/c)^2$, that is $<10^{-5}$ in shocks with $v_s<1000$~km~s$^{-1}$, compared to an energy released per unit mass of  $\sim 0.1$ for accretion into a BH \citep[see discussion in][]{laor1998}. 

        Shocks may be produced by an AGN driven wind. Can the AGN drive the required massive gas outflow? The implied  momentum flux in the shock is 
        $\dot{P_s}=\dot{M_s} v_s$, while the AGN radiative momentum flux is
        $\tau L_{\rm bol}/c$, where $\tau$ is the flux weighted mean optical depth of the wind. The $\dot{M_s}$ values derived above imply 
         \begin{equation}\label{Pdot}
              \dot{P_s}= (284, 67, 21)\times L_{\rm bol}/c\ .        
        \end{equation}
        The required range in $\tau=21-284$ may be possible only in highly obscured AGN. It is not possible for typical AGN winds to create the very high momentum flux needed through the shocks. 
       
        The implied kinetic energy flux in the shocked gas is
        $\dot{E_s}=\dot{M_s} v_s^2/2$, which is
         \begin{equation}\label{Edot}
              \dot{E_s}= (0.095, 0.067, 0.035)\times L_{\rm bol}         
        \end{equation}
         for the above $v_s$. Such a high efficiency in the conversion of the AGN $L_{\rm bol}$ to winds kinetic energy is generally not observed, and is also not derived in simulations. 
         
         Shocks therefore require rather extreme $\dot{M_s}, \dot{P_s}$ and
         $\dot{E_s}$ values to produce the observed CL luminosities.  A massive highly optically thick outflow may be relevant in gas rich ULIRG systems, but is unlikely to be significant in the more common unobscured AGN. 
         Shocks may be more relevant in LLAGN, where the line emission is powered mostly by the host stellar processes. If significant AGN luminosity is present, that is at the level it can power most of the NLR emission, then it will most likely also power the CL emission.

\section{Summary and Conclusions} \label{sec:conclusions}
    
    In this paper we have conducted \textsc{Cloudy} calculations of various slabs of gas confined by radiation pressure. 
    The gas is either with or without dust, with either supersolar or subsolar metallicity, and with the range of
    observed ionizing SEDs at $L_{\mathrm{bol}} = 10^{43}-10^{47}\,$erg\,s$^{-1}$. 
    We derive the conditions which are optimal for detecting optical CLs. 
    Our main conclusions are:
    \begin{itemize}
        \item CLs are heavily suppressed in dusty gas, by factors of $10-10^3$ in luminosity and equivalent width 
        (Fig.~\ref{fig1}), and will generally be non detectable
        (EW$<0.1$\AA) by current large surveys (e.g., SDSS). This is due to CL emitting species such as calcium and iron being heavily depleted onto grains, and grains absorbing ionizing photons and thus shrinking the photoionized gas volume by a factor of $\sim$10. In dustless gas, most of the CLs studied here (Table \ref{linetable}) are expected to be detectable. 
        \item The production of optical CLs does not require a separate gas component of highly ionized gas, or an exceptionally hard ionizing SED (Fig.\ref{fig3}). CLs can be produced by all gas clouds from sub pc to kpc scales. Since the density of RPC gas scales as $r^{-2}$, the CL emission is stratified with $r$ with peak emission at the $0.1-10\,$pc range, compared to $10-100\,$pc range for lines from less highly ionized species, which are characterized by a lower critical density. 
        \item The [Ne\,V] 3425.88\,\AA\ line is the brightest CL in both dusty and dustless gas. It decreases in luminosity by a factor of $\lesssim$10  in dusty gas, as Neon is not depleted to grains. It is thus likely to be detectable in a significant fraction of bright AGNs.
        \item A hard ionizing SED enhances some of the CLs by a factor of $\sim 10$, but all CLs (excluding [Ne\,V] 3425.88\,\AA) still remain non detectable if the gas is dusty (Fig.\ref{fig5}). More prominent CLs are expected in objects with dustless gas and a hard ionizing SED.          
        \item We suggest that the optical CLs in dwarf galaxies may result from their unusually low metallicity and the associated low dust content and smaller depletion to grains. The lower black hole masses expected in dwarf galaxies suggests a harder ionizing SED, which together with their low $Z$ is expected to enhance their CL emission. 
        \item The [O\, III] 5007\,\AA\ line is the strongest lower IP line, and remains detectable for dusty gas (see Fig.\ref{fig2}).
        It becomes exceptionally strong (EW$>100$\AA) for dustless gas and a hard continuum, which may thus serve as a useful tracer for CL emission in AGN \citep{Thomas2017}. In contrast, the lower ionization lines of [N II], [S II] and [O I], become more prominent in dusty gas and their prominence may argue against the presence of optical CLs.
        \item Previous studies have suggested shocks as a means to destroy dust grains in the ISM (e,g, \citealp{1998ApJ...501..643D, 2004ASPC..309..347J, 2015ApJ...803....7S}) and AGN outflows have been shown to have low reddening values (e.g., \citealp{2003ARA&A..41..117C, 2014MNRAS.445.3011S}). CLs in AGNs and dwarf galaxies have also been shown to be associated with outflows (e.g., \citealp{2021ApJ...911...70B}; Matzko et al.\,\,submitted; Doan et al.\,\,submitted). Thus, although shocks are likely too weak to produce significant line emission in AGN,  shocks may be necessary as a grain destruction mechanism which allow for photoionization to produce significant CLs.  
    \end{itemize}

In this work, we have demonstrated that optical CLs are faint in the spectra of AGNs where the gas in the NLR 
is dusty. Which mechanism reduces the dust content of the gas in AGN with strong CL remains to be explored.

\begin{acknowledgments}

The Authors thank Gary Ferland and the entire \textsc{Cloudy} team for developing and maintaining \textsc{Cloudy}.

J.D.M would like to acknowledge Joseph C. Weingartner for the many useful conversations regarding the physics contained in this manuscript, and Jonathan Stern for advice on using \textsc{Cloudy} in the context of RPC. J.D.M also acknowledges the Doctoral Research Scholars grant from the office of the provost of George Mason University, VA. C.R acknowledges support from Fondecyt Regular grant 1230345 and ANID BASAL project FB210003. J.M.C.'s research is supported by an appointment to the NASA Postdoctoral Program at the NASA Goddard Space, administered by Oak Ridge Associated Universities under contract with NASA. The Cloudy calculations were run on HOPPER, a research computing cluster provided by the Office of Research Computing at George Mason University, VA. (http://orc.gmu.edu) 

\end{acknowledgments}

\bibliography{bibi}{}
\bibliographystyle{aasjournal}

\end{document}